\documentclass[]{aa}  
\usepackage{graphicx}
\usepackage[varg]{txfonts}
\usepackage{longtable}
\usepackage{threeparttable}

\newcommand{\new}[1]{{#1}}

\begin{document} 

\title{Erosion and the limits to planetesimal growth}
\subtitle{}

   \author{S. Krijt\inst{1}
          \and
          C.W. Ormel\inst{2,3}
          \and
          C. Dominik\inst{3}
          \and
          A.G.G.M. Tielens\inst{1}}
      
\institute{Leiden Observatory, Leiden University, Niels Bohrweg 2, 2333 CA Leiden, The Netherlands\\
              \email{krijt@strw.leidenuniv.nl}\label{inst1}
\and Astronomy Department, University of California, Berkeley, CA 94720\label{inst2}
\and Anton Pannekoek Institute, University of Amsterdam, Science Park 904, 1098 XH Amsterdam, The Netherlands\label{inst3}}

   \date{}

 \abstract
 {The coagulation of microscopic dust into planetesimals is the first step towards the formation of planets. The composition, size, and shape of the growing aggregates determine the efficiency of this early growth. In particular, it has been proposed that fluffy ice aggregates can grow very efficiently in protoplanetary disks, suffering less from the bouncing and radial drift barriers.}
 {While the collision velocity between icy aggregates of similar size is thought to stay below the fragmentation threshold, they may nonetheless lose mass from collisions with much smaller projectiles. As a result,  erosive collisions have the potential to terminate the growth of pre-planetesimal bodies. We investigate the effect of these erosive collisions on the ability of porous ice aggregates to cross the radial drift barrier.}
 {We develop a Monte Carlo code that calculates the evolution of the masses and porosities of growing aggregates, while resolving the entire mass distribution at all times. The aggregate's porosity is treated independently of its mass, and is determined by collisional compaction, gas compaction, and eventually self-gravity compaction. We include erosive collisions and study the effect of the erosion threshold velocity on aggregate growth.}
 {For erosion threshold velocities of $20-40\mathrm{~m~s^{-1}}$, high-velocity collisions with small projectiles prevent the largest aggregates from growing when they start to drift. \new{In these cases, our local simulations result in a steady-state distribution, with the majority of the dust mass in particles with Stokes numbers close to unity. Only for the highest erosion threshold considered ($60\mathrm{~m~s^{-1}}$), do porous aggregates manage to cross the radial drift barrier in the inner 10 AU of MMSN-like disks.}}
 {Erosive collisions are more effective in limiting the growth than fragmentary collisions between similar-size particles. Conceivably, erosion limits the growth before the radial drift barrier, although the robustness of this statement depends on (uncertain) material properties of icy aggregates. If erosion inhibits planetesimal formation through direct sticking, the sea of ${\sim}10^9$ g, highly porous particles \new{appears well-suited for triggering streaming instability}.}
  
\keywords{protoplanetary disks - planet and satellites: formation - stars: circumstellar matter - methods: numerical}

\maketitle

\section{Introduction}
Despite the apparent ease with which nature is forming planets, current models of planet and even planetesimal formation have problems growing large bodies within the typical gas disk lifetime of ${\sim}10^{6}$ years \citep{haisch2001}. The process of planetesimal formation is a complex one, with many different processes acting on a variety of length and timescales (see \citet{testi2014} and \citet{Johansen2014} for recent reviews).

The first step towards planetesimal formation is the coagulation of small dust aggregates that stick together through surface forces. As aggregates collide and stick to form larger aggregates, these aggregates have to overcome several hurdles on their way to becoming planetesimals. One important obstacle faced by a growing dust aggregate is the radial drift barrier \citep{whipple1972, weidenschilling1977}. When aggregates grow to a certain size (about a meter at 1 AU and a millimeter at 100 AU, assuming compact particles) they will decouple from the pressure-supported gas disk, and start to lose angular momentum to the gas around them. As a result, said particles will drift inward.

But even before radial drift becomes problematic, coagulation of aggregates can be frustrated by catastrophic fragmentation or bouncing \citep{blumwurm2008, guttler2010, zsom2010}, which prevent colliding aggregates from gaining mass. These issues are alleviated somewhat by including velocity distributions between pairs of particles \citep{windmark2012b, garaud2013} in combination with mass transfer in high-velocity collisions \citep{wurm2005,kothe2010}, though these solutions require the presence of \new{relatively} compact targets. 

Recently, it was proposed that icy aggregates, if they can manage to stay very porous, suffer less from these barriers, and might be able to form planetesimals locally and on relatively short timescales \citep{okuzumi2012, kataoka2013c}. Very porous, or fluffy, aggregates are less likely to bounce \citep{wada2011,seizinger2013a}, and icy particles have much higher fragmentation threshold velocities than refractory ones \citep{dominiktielens1997,wada2013}. But perhaps most surprising was the finding that porous aggregates can outgrow the radial drift barrier, by growing very rapidly due to their enhanced collisional cross section \citep{okuzumi2012}. However, \citet{okuzumi2012} assumed perfect sticking between colliding aggregates, \new{neglecting possible mass-loss in aggregate-aggregate collisions.}

\new{In this paper, we study the effects of the existence of an erosive regime for icy aggregates, where collisions at low mass ratios will produce erosive fragments at velocities below a critical erosion threshold velocity \citep{schrapler2011, seizinger2013c, gundlach2014}.} Our goal is to quantify how erosion influences the direct formation of planetesimals through coagulation. To this end, we develop a local Monte Carlo coagulation code, capable of simulating the vertically-integrated dust population, tracing both the evolution of the mass and the porosity of the entire mass distribution self-consistently. Section \ref{sec:diskdust} describes the models we use for the protoplanetary disk and the dust aggregates. In section \ref{sec:MC}, we present the numerical method, which is based on the work of \citet{ormel2008}. Then, we test our model against the results of \citet{okuzumi2012} (Section \ref{sec:perfect_sticking_a}), after which we expand the model to include compaction from gas pressure and self-gravity according to \citet{kataoka2013c} (Section \ref{sec:perfect_sticking_b}), and erosive collisions (Section \ref{sec:erosion_sims}). In Section \ref{sec:semian}, we compare the results to a simple semi-analytical model, and describe which processes can limit coagulation in different parts of protoplanetary disks. \new{Discussion of the results and implications takes place in Section \ref{sec:discussion}, and conclusions are presented in Section \ref{sec:conclusions}.}

\section{Disk and dust models}\label{sec:diskdust}
The disk model and collisional compaction prescription are based on \citet{okuzumi2012}, to which we add non-collisional compaction processes (Section \ref{sec:gasgrav}) and a model for erosive collisions \new{(Sections \ref{sec:erosion_case} and \ref{sec:erosion_model})}. 

\subsection{Disk structure}\label{sec:disk}
The disk model used in this work is based on the minimum-mass solar nebula (MMSN) of \citet{hayashi1981}. The evolution of the gas surface density and temperature as a function of radial distance $R$ from the Sun-like central star are given as
\begin{equation}
\Sigma_g = 152 \left(\frac{R}{5\mathrm{~AU}}\right)^{-3/2} \mathrm{~g~cm^{-2}},
\end{equation}
\begin{equation}
T = 125 \left(\frac{R}{5\mathrm{~AU}}\right)^{-1/2} \mathrm{~K}.
\end{equation}
The gas sound speed is given by
\begin{equation}
c_s = \sqrt{k_{\mathrm{B}} T / m_g} =6.7\times10^{2} \left(\frac{R}{5\mathrm{~AU}}\right)^{-1/4} \mathrm{~m~s^{-1}},
\end{equation}
with $k_{\mathrm B}$ the Boltzmann constant and $m_g = 3.9\times10^{-24}\mathrm{~g}$ the mean molecular weight. The Kepler frequency equals
\begin{equation}
\Omega = \sqrt{GM_{\odot}/R^3} = 1.8\times10^{-8} \left(\frac{R}{5\mathrm{~AU}}\right)^{-3/2} \mathrm{~s^{-1}}.
\end{equation} 
Assuming an isothermal column, the gas density drops with increasing distance from the mid plane $z$ according to
\begin{equation}\label{eq:rho_g}
\rho_g = \frac{\Sigma_g}{\sqrt{2\pi}h_g}\exp\left(\frac{-z^2}{2h_g^2} \right),
\end{equation}
with the relative vertical scale height of the gas $h_g/R=0.05(R/\mathrm{5~AU})^{1/4}$. The turbulent viscosity is parametrized as $\nu_\mathrm{turb} = \alpha c_s^2  / \Omega$ following \citet{shakura1973}, and $\alpha$ is assumed to be constant in both the radial and the vertical direction. The eddie turn-over time of the largest eddies equals $t_L=\Omega^{-1}$.

In our local model, the surface density of the dust is related to the gas surface density through $\Sigma_d/\Sigma_g = 10^{-2}$, but the vertical distribution of dust depends on its aerodynamic properties. The dust is described by a Gaussian, with the dust scale height $h_d$ set by the stopping time $t_s$ of the dust particle through \new{\citep{youdin2007}}
\begin{equation}\label{eq:h_d}
\frac{h_d}{h_g} = \left(1+ \frac{\Omega t_s}{\alpha}\frac{1+2\Omega t_s}{1+\Omega t_s} \right)^{-1/2}.
\end{equation}
Thus, settling becomes important when a dust particle reaches $\Omega t_s \sim \alpha$.

\subsection{Dust properties}
Initially, all dust particles are assumed to be spherical (sub)micron-size monomers. In time, these monomers coagulate through collisions, and aggregates of considerable mass can be formed. Any aggregate is described by two parameters: the mass $m$, and the filling factor $\phi$. Since aggregates are made up of monomers the mass can be written as $m=Nm_0$, with $N$ the number of monomers and $m_0$ the monomer mass. Following \citet{okuzumi2012}, we define the internal density of an aggregate as $\rho_{\mathrm{int}} = m/V$, with $V=(4/3)\pi a^3$ the volume of the aggregate, and $a$ its radius. An aggregate's radius is defined as $a=[5/(3N) \sum_{k=1}^N (\vec{r}_k - \vec{r}_\mathrm{CM})^2 ]^{1/2}$, with $\vec{r}_k$ the position of monomer $k$ and $\vec{r}_\mathrm{CM}$ the position of the aggregate's center of mass \citep{mukai1992,suyama2008,okuzumi2009}. By definition, monomers have an internal density of $\rho_{\mathrm{int}} = m_0/V_0 = \rho_0$, while aggregates can have $\rho_{\mathrm{int}}\ll \rho_0$. Since we are interested in region beyond the snow-line, we focus here on monomers composed of mostly ice, and use a density of $\rho_0 = 1.4 \mathrm{~g~cm^{-3}}$. For the monomer radius we use $a_0=0.1\mathrm{~\mu m}$. We define the filling factor as 
\begin{equation}
\phi \equiv \frac{\rho_{\mathrm{int}}}{\rho_0},
\end{equation}
as a measure for the internal density.

In the rest of this section, we describe the main ingredients for the simulations presented in Section \ref{sec:MC}. These are: the relative velocities between aggregates, the equations governing the evolution of $\rho_{\mathrm{int}}$ through mutual collisions as well as gas ram pressure and self-gravity, and models for the destructive processes of erosion and fragmentation.

\subsubsection{Relative velocities}
We take into account relative velocities arising from Brownian motion, turbulence, settling, radial drift and azimuthal drift \citep[see Section 2.3.2 of][]{okuzumi2012}. The relative contribution of the velocity components depends strongly on the size and aerodynamic properties of the dust grains in question. More specifically, the relative velocity is a function of the stopping times of the particles. Depending on the size of the particle, the stopping time is set either by Epstein or Stokes drag
\begin{equation}\label{eq:t_s}
t_s = \begin{cases} 
t_s^{\mathrm{(Ep)}} = \dfrac{3m}{4\rho_g v_{\mathrm{th}}A}  &\textrm{~~for~~} a < \dfrac{9}{4}\lambda_{\mathrm{mfp}},\vspace{2mm} \\

t_s^{\mathrm{(St)}} = \dfrac{4a }{9 \lambda_{\mathrm{mfp}}}t_s^{\mathrm{(Ep)}}  &\textrm{~~for~~} a > \dfrac{9}{4}\lambda_{\mathrm{mfp}},
\end{cases}
\end{equation}
where $v_{\mathrm{th}} = \sqrt{8/\pi}c_s$ is the mean thermal velocity of the gas molecules, and $\lambda_{\mathrm{mfp}} = m_g/(\sigma_{\mathrm{mol}}\rho_g)$ is the gas molecule mean free path. Taking $\sigma_{\mathrm{mol}}=2\times10^{-15}\mathrm{~cm^{2}}$, we obtain $\lambda_{\mathrm{mfp}} = 120(R/\mathrm{5~AU})^{11/4}\mathrm{~cm}$ at the disk mid plane. In Equation \ref{eq:t_s}, $a=a_0 (V/V_0)^{1/3}$ refers to the dust particle radius, while $A$ is the projected cross section of the particle averaged over all orientations, which can be obtained using the formulation of \citet{okuzumi2009}.

\begin{figure}
\includegraphics[clip=,width=1.\linewidth]{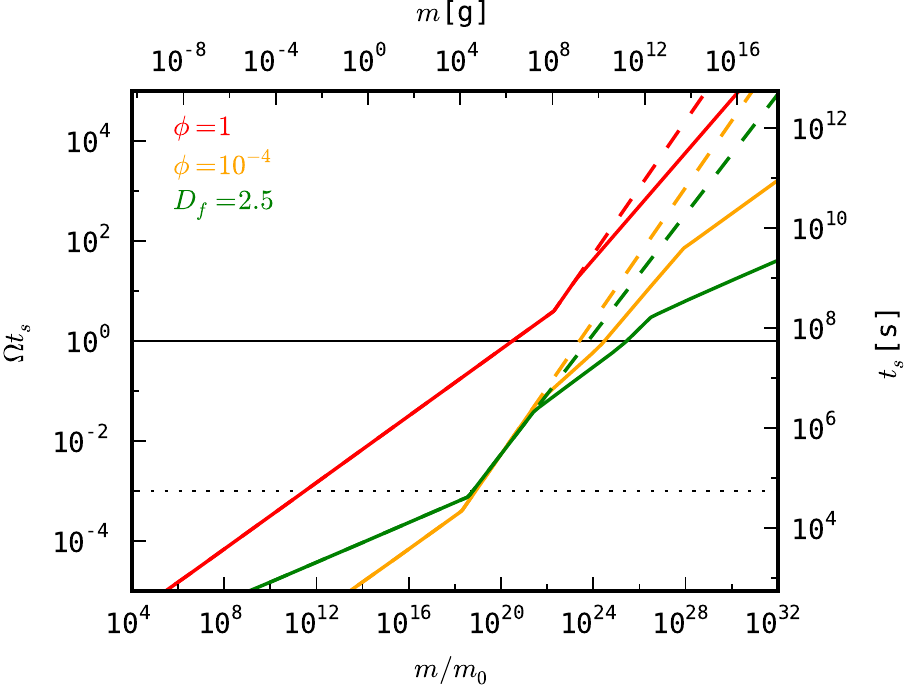}
\caption{Particle Stokes numbers as a function of mass, in the mid plane of an MMSN disk at 5 AU. Different lines show compact particles (red), porous aggregates with constant $\phi=10^4$ (yellow), and aggregates with a constant fractal dimension of 2.5 (green). For the solid lines, all drag regimes (Epstein, Stokes and Newton) have been taken into account, while the dashed lines indicate the results using Epstein and Stokes drag only. Horizontal lines indicate $\Omega t_s=1$ (where drift is fastest) and $\Omega t_s=\alpha=10^{-3}$ (where particles start to settle to the mid plane).}
\label{fig:stokes} 
\end{figure}

The above equation is accurate when the particle Reynolds number $\mathrm{Re_p}= 4 a v_\mathrm{dg}/(v_\mathrm{th} \lambda_\mathrm{mfp}) <1$, with $v_\mathrm{dg}$ the relative velocity between the gas and the dust particle. The Reynolds number can become large when aggregates grow very big or their velocity relative to the gas is very large. In general, the stopping time can be written as
\begin{equation}
t_s = \frac{2m}{C_D \rho_g v_\mathrm{dg} A}.
\end{equation}
In the Stokes regime the drag coefficient equals $C_D=24/\mathrm{Re_p}$, and the stopping time becomes independent of $v_\mathrm{dg}$. However, for larger Reynolds number the stopping time becomes a function of the velocity relative to the gas. This regime is called the Newton drag regime. Since the relative velocity depends in turn on the stopping time, we have to iterate to find the corresponding stopping time. Following \citet{weidenschilling1977}, we use
\begin{equation}\label{eq:newton}
C_D = \begin{cases}
~~24(\mathrm{Re_p})^{-1} &\textrm{~~for~~} \mathrm{Re_p}<1,\vspace{2mm} \\
~~24(\mathrm{Re_p})^{-3/5} &\textrm{~~for~~} 1<\mathrm{Re_p}<800,\vspace{2mm} \\
~~0.44&\textrm{~~for~~} 800<\mathrm{Re_p}.
\end{cases}
\end{equation}
Figure \ref{fig:stokes} shows Stokes numbers ($\Omega t_s$) for different particles in the mid plane of a MMSN disk at 5 AU. Different lines show compact particles (red), porous aggregates with constant $\phi=10^4$ (yellow), and aggregates with a constant fractal dimension of 2.5 (green). For the solid lines, all drag regimes (Epstein, Stokes and Newton) have been taken into account, while the dashed lines indicate the results using Epstein and Stokes drag only, i.e. assuming that $\mathrm{Re_p}<1$. Focussing on the $D_f=2.5$ aggregates, we can clearly distinguish the different drag regimes. The smallest particles are in the Epstein regime, and switch to the Stokes regime around $\Omega t_s=10^{-3}$. Then, at a mass of $m/m_0 \sim 10^{21}$, the Reynolds number exceeds unity and we enter the second regime of Equation \ref{eq:newton}. Note that this transition occurs before $\Omega t_s =1$. The most massive particles, $m/m_0>10^{26}$ are in the regime where $C_D=0.44$. Compact particles on the other hand, reach $\Omega t_s=1$ while still in the Epstein drag regime.

The turbulence-induced relative velocity between two particles with stopping times $t_{s,1}$ and $t_{s,2}\leq t_{s,1}$ has three regimes \citep{ormel2007b}
\begin{equation}\label{eq:v_turb} 
v_\mathrm{turb} \simeq \delta v_g \times \begin{cases}
~~\mathrm{Re_t}^{1/4} \ \Omega (t_{s,1}-t_{s,2}) &\textrm{~for~} t_{s,1} \ll t_\eta,\vspace{2mm} \\
~~1.4\dots1.7 \left(\Omega t_{s,1} \right)^{1/2} &\textrm{~for~} t_\eta \ll t_{s,1} \ll \Omega^{-1}, \vspace{1mm} \\ 
~~\left(\dfrac{1}{1+\Omega t_{s,1}}  + \dfrac{1}{1+\Omega t_{s,2}} \right)^{1/2}&\textrm{~for~} t_{s,1}\gg \Omega^{-1},
\end{cases}
\end{equation}
where $\delta v_g = \alpha^{1/2} c_s$ is the mean random velocity of the largest turbulent eddies, and $t_\eta= \mathrm{Re_t}^{1/2} t_L$ is the turn-over time of the smallest eddies. The turbulence Reynolds number is given by $\mathrm{Re_t}= \alpha c_s^2 / (\Omega \nu_\mathrm{mol} )$, with the molecular viscosity $\nu_\mathrm{mol}=v_\mathrm{th} \lambda_\mathrm{mfp}/2$. We will refer to the first two cases of Equation \ref{eq:v_turb} as the first and second turbulence regimes. Relative velocities between similar particles (similar in the sense that they have comparable stopping times) are very small\footnote{According to Equation \ref{eq:v_turb}, $v_\mathrm{turb}=0$ for aggregates with identical stopping times in the first turbulence regime. In reality, the dispersion in the aggregate's mass-to-area ratio will give rise to a small relative velocity. We treat this dispersion in the same way as \citet{okuzumi2012}, by taking into account the standard deviation in the mass-to-area ratio of a porous aggregate \citep{okuzumi2011}. The size of this standard deviation, normalized by the mean mass-to-area ratio, is parametrized as $\varepsilon$, which we take to equal 0.1, following \citet{okuzumi2011}.} in the first turbulence regime because of the $(t_{s,1}-t_{s,2})$ term, but considerably larger in the second regime. 

\begin{figure*}
\includegraphics[clip=,width=1.\linewidth]{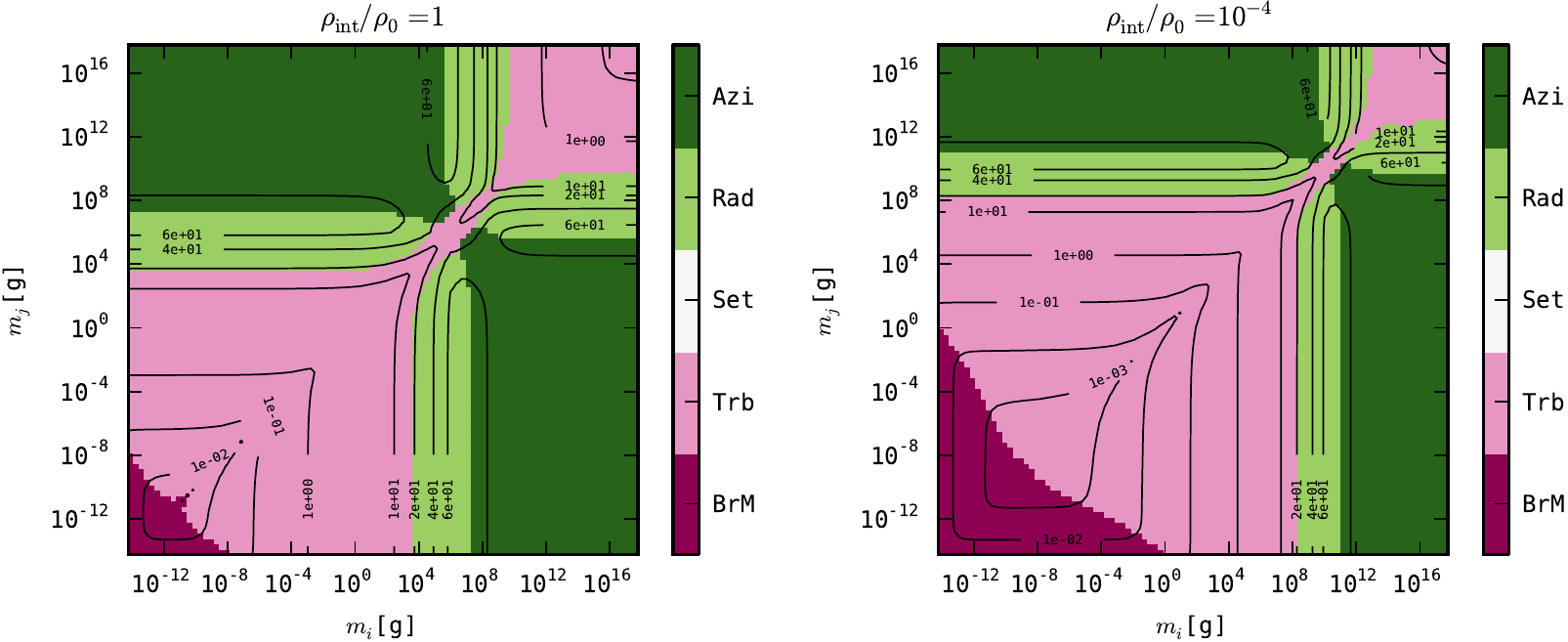}
\caption{Relative velocities between compact (left) or very porous (right) particles with masses $m_i$ and $m_j$ at the mid plane of an MMSN-disk at 5 AU with $\alpha=10^{-3}$. The masses range from single monomers to aggregates containing $10^{32}$ monomers. The contours give relative velocities in $\mathrm{m~s^{-1}}$, and the colors indicate the dominating source for the relative velocity: Brownian motion (BrM), turbulence (Trb), settling (Set), radial drift (Rad), or azimuthal drift (Azi). Epstein, Stokes, and Newton drag regimes have been taken into account.}
\label{fig:velocities}
\end{figure*}

Figure \ref{fig:velocities} shows the mid plane relative velocity in $\mathrm{m~s^{-1}}$ (contours), and its dominant source (color), for a range of combinations of masses $m_i$ and $m_j$. The velocities have been calculated for the disk properties of Section \ref{sec:disk}, at 5 AU, and assuming a turbulence $\alpha=10^{-3}$. The left plot corresponds to two compact particles ($\rho_{\mathrm{int}}=\rho_0$), and the right plot to two very porous ones ($\rho_{\mathrm{int}}=10^{-4}\rho_0$). The general picture is the same for all porosities: Brownian motion dominates the relative velocity at the smallest sizes, followed by turbulence for larger particles, and systematic drift for bodies that have $\Omega t_s \sim 1$. However, the masses at which various transitions occur can vary orders of magnitude depending on the particle porosity. \new{For this particular location and turbulence strength, there is no combination of particle masses whose relative velocity is dominated by differential settling.}

\subsection{Collisional outcomes}
A collision between porous aggregates can have a number of outcomes, ranging from perfect sticking to catastrophic fragmentation. For silicates, \citet{blumwurm2008} and \citet{guttler2010} offer reviews of the various outcomes as observed in laboratory experiments. For porous ices, experimental investigations are scarce, and we have to turn to numerical simulations when predicting the outcome \citep[e.g.][]{dominiktielens1997,wada2007,suyama2008,wada2009}. 

In general, a collision can result in sticking, erosion, or fragmentation, depending on the relative velocity and the mass ratio $R^{(m)} \equiv m_i/m_j \leq 1$ of the colliding bodies. Collisions between particles with comparable masses result in catastrophic fragmentation if they collide above the fragmentation velocity (Section \ref{sec:fragmentation}). When colliding bodies have a mass ratio $R^{(m)} \ll 1$, catastrophic fragmentation of the larger body is difficult, but the collision can result in erosion if the velocity is high enough. The transition from erosion to the fragmentation regime occurs at a mass ratio $R^{(m)}_\mathrm{crit}$, specified in Section \ref{sec:erosion_model}. In an erosive event, the larger body will lose mass. From Figure \ref{fig:velocities} it is clear that the highest velocities are reached between particles with very different masses, and thus erosion might very well be a common collisional outcome. We discuss erosion in more detail in Section \ref{sec:erosion_case}. We should note at this point that we do not consider bouncing collisions. For relatively compact silicate particles, bouncing is frequently observed in the laboratory \citep[e.g.][]{guttler2010}, and indeed can halt growth in protoplanetary disks \citep{zsom2010}. However, in porous aggregates, the average coordination number (the number of contacts per monomer) is much lower than in compact ones. As a result, collision energy is more easily dissipated, and it is safe to neglect bouncing \citep{wada2011,seizinger2013a}.

\subsubsection{Catastrophic fragmentation}\label{sec:fragmentation}
For collisions between roughly equal icy aggregates (mass ratio $R^{(m)} \geq 1/64$), \citet{wada2013} find a critical fragmentation velocity of 
\begin{equation}\label{eq:v_frag}
v_{\mathrm{frag}} \simeq 20 \left(\frac{E_\mathrm{break}}{m_0}\right)^{1/2} \simeq 80 \left(\frac{a_0}{0.1\mathrm{~\mu m}}\right)^{-5/6} \mathrm{~m~s^{-1}}.
\end{equation}
The quantity $E_\mathrm{break}$ represents the energy needed to break a single mononer-monomer contact \citep{dominiktielens1997}. Collisions below this critical velocity result in sticking, while collisions at or above $v_{\mathrm{frag}}$ result in fragmentation of the collision partners.


\begin{figure}
\includegraphics[clip=,width=.9\linewidth]{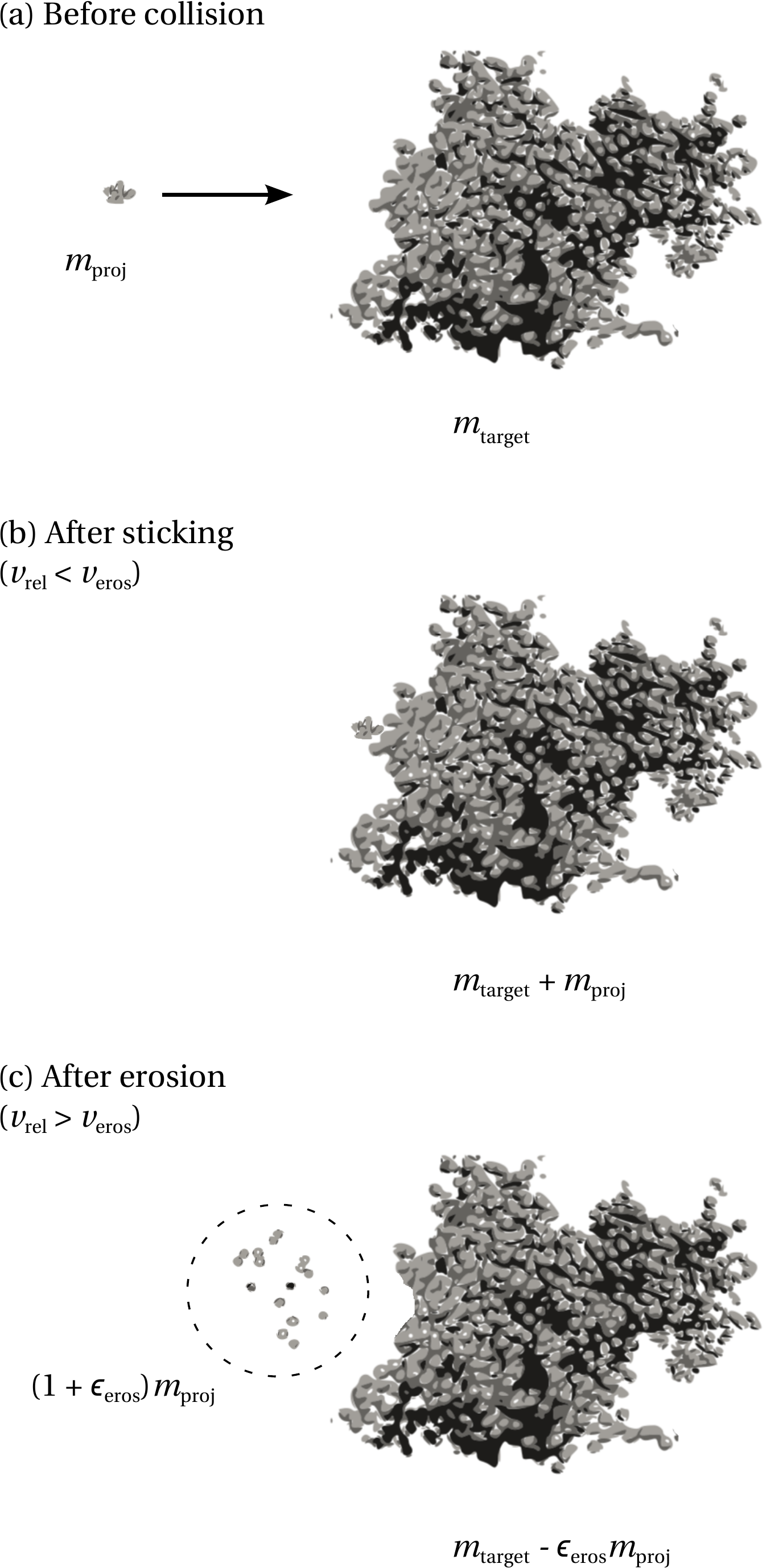}
\caption{\emph{(a):} Schematic of a collision between unequal particles with a mass ratio $R^{(m)} = (m_\mathrm{proj}/m_\mathrm{target})\ll1$. \emph{(b):} Sticking occurs when $v_\mathrm{rel}<v_\mathrm{eros}$. The mass of the projectile is added to the target. \emph{(c):} Collisions above the erosion threshold velocity lead to erosion. The mass loss of the target is given by the erosion efficiency $\epsilon_\mathrm{eros}$ and the mass of the projectile.}
\label{fig:schem}
\end{figure}

\subsubsection{The case for erosion}\label{sec:erosion_case}
\new{The relative velocity between similar-sized aggregates will generally not reach the fragmentation velocity (Equation \ref{eq:v_frag}) behind the snow line in a protoplanetary disk, especially not if the turbulence is weak. However, relative velocities between particles with very different masses can be much larger than velocities between similar particles, especially when radial and azimuthal drift are important (Figure \ref{fig:velocities}). In this paper, we study the effects of an erosive regime, where collisions at low mass ratios will produce erosive fragments at velocities below a critical erosion threshold velocity $v_\mathrm{eros} \lesssim v_\mathrm{frag}$. Here, we briefly revisit numerical and experimental studies of erosion, before outlining the erosion model used in this work. The process of erosion can be described by two main quantities: the erosion threshold velocity, $v_{\mathrm{eros}}$, above which erosion takes place, and the (normalized) erosion efficiency, $\epsilon_{\mathrm{eros}}$, that indicates how much mass is eroded in units of projectile mass.}

\new{For silicate particles, \citet{guttler2010} summarize a number of experimental investigations and describe a threshold velocity of a few $\mathrm{m~s^{-1}}$, and an erosion efficiency that increases roughly linearly with collision velocity. Similar trends were observed by \citet{schrapler2011}, who found an erosion threshold velocity of a few $\mathrm{m~s^{-1}}$ using micron-size silicate projectiles. Note that the threshold velocity is comparable to the monomer sticking velocity of micron-size silicate particles \citep{poppe2000}. In the experiments of \citet{schrapler2011}, the erosion efficiency also increased with impact velocity, reaching ${\sim}10$ for the highest velocity of $60\mathrm{~m~s^{-1}}$ (their Figure 5). \citet{seizinger2013c} used molecular dynamics simulations, based on a new viscoelastic model \citep{krijt2013}, to reproduce the experimental results. In addition, \citet{seizinger2013c} studied the variation on the threshold velocity and erosion efficiency with projectile mass, showing a trend of decreasing erosion threshold with decreasing mass ratio (e.g. their Figure 11). For monomer projectiles, the threshold velocity equals the monomer-monomer sticking velocity $v_s \simeq \sqrt{E_\mathrm{break}/m_0}$, after which it increased linearly with velocity to eventually flatten off around $10\mathrm{~m~s^{-1}}$. This flattening off indicates the onset of catastrophic fragmentation, and occurs at a mass ratio of ${\sim}10^{-2}$.}

\new{For ice particles, \citet{gundlach2014} present recent experimental results on the sticking and erosion threshold of (sub)micron-size particles. For a projectile distribution between $0.2-6\mathrm{~\mu m}$ (with a mean value of $1.5\mathrm{~\mu m}$) impinging an icy target with a filling factor $\phi\simeq0.5$, an erosion threshold of $15.3\mathrm{~m~s^{-1}}$ was found. These results confirm the increased stickiness of ice compared to silicate particles, and indicate $v_\mathrm{eros}$ could indeed be very high for (monodisperse) 0.1-$\mathrm{\mu m}$ monomers, possibly even ${>}60\mathrm{~m~s^{-1}}$. However, the aggregates acting as targets in the simulations presented here have a much higher porosity ($\phi \sim 10^{-3}$), and the lower coordination number is expected to reduce the erosion threshold \citep{dominiktielens1997}. Lastly, while \citet{gundlach2014} used a distribution of grain sizes, numerical investigations \citep[e.g.,][]{seizinger2013c,wada2013}, for computational reasons, often employ a monodisperse monomer distribution, making a direct comparison difficult. For a single grain size, the size significantly influences the strength of the aggregates, with larger monomers leading to weaker aggregates (Equation \ref{eq:v_frag}). Little is known about the expected grain sizes in the icy regions of protoplanetary disks, let alone their size distribution, or about the effect a monomer size distribution has on the strength and collisional behavior of porous aggregates.}

\new{For these reasons, we believe that the existence of an erosive regime for icy aggregates is plausible. However, at present the data are unfortunately ambiguous with other simulations indicating the opposite trend: that the mass-loss in low mass ratio collisions is relatively small. Using molecular dynamics N-body simulations \citet{wada2013} find that the threshold velocity (where fragmentary collisions become more numerous than sticky collisions) increases for smaller mass ratios, suggesting that only similar-size particles colliding at $v_\mathrm{frag}$  fragment efficiently. This trend of an increased erosion threshold for smaller size ratios is corroborated by recent simulations by Tanaka et al. (in prep). This would imply that for monodisperse submicron grains, both threshold velocities might not be reached (cf. Equation \ref{eq:v_frag} and Figure \ref{fig:velocities}). In this paper, we take the agnostic view by `burying' the uncertainty of the erosion threshold velocity in the parameter $v_\mathrm{eros}$, which we vary to investigate the implications of effective versus ineffective erosion.}

\subsubsection{Erosion model}\label{sec:erosion_model}
\new{Erosive collisions occur only below a mass ratio $R^{(m)}_\mathrm{crit}$, and their outcome is parametrized in terms of a (velocity-dependent) erosion efficiency. In accordance with \citet{guttler2010} and \citet{seizinger2013c} we will use $R^{(m)}_\mathrm{crit}=10^{-2}$. For smaller mass ratios, we will assume a constant value for $v_\mathrm{eros}$, that does not depend on mass ratio or projectile/target porosity. We vary $v_\mathrm{eros}$ between 20 and $60\mathrm{~m~s^{-1}}$, corresponding to $(1/4)v_\mathrm{frag}$ and $(3/4)v_\mathrm{frag}$ for $0.1\mathrm{~\mu m}$ monomers (Equation \ref{eq:v_frag}). In the erosive regime, the normalized erosion efficiency can be written as}
\begin{equation}\label{eq:e_eros}
\epsilon_\mathrm{eros} = c_1 \left( \frac{v_{\mathrm{rel}}}{v_{\mathrm{eros}}} \right)^{\gamma},
\end{equation}
with $c_1\sim1$ \citep{guttler2010,seizinger2013c}. While in supersonic cratering collisions $\gamma=16/9$ \citep{tielens1994}, the velocities encountered in this work are not that high and at most comparable to the sound speed in porous aggregates \citep{paszun2008}. Hence, we will use $\gamma=1$, in agreement with both numerical and experimental work in the appropriate velocity range \citep{guttler2010, schrapler2011, seizinger2013c}.

Lastly, we need a prescription for the filling factors after an erosive collision. We assume that $i)$ the filling factor of the target remains unchanged, and $ii)$ the filling factor of the fragments is found by assuming they have the same fractal dimension as the target, where the target's fractal dimension $D_f$ is estimated as
\begin{equation}
D_f \simeq 3 \left[1- \frac{\log(\phi)}{\log(m/m_0)} \right]^{-1}.
\end{equation}
\new{The assumptions of the erosion model employed in this work are discussed further in Section \ref{sec:discussion}.}

\subsection{Aggregate compaction}
\new{An aggregate's porosity can be altered through collisions, or through non-collisional mechanisms. In this Section, we first describe how porosity can increase and decrease as the result of sticking collisions. Then, we discuss gas- and self-gravity compaction.}

\subsubsection{Collisional compaction}
When two particles $i$ and $j$ collide at a relative velocity $v_{\mathrm{rel}}$ that is below the thresholds for fragmentation or erosion, the particles stick, and form a new aggregate with mass $m_i+m_j$. The internal density of the new particle depends on how the impact energy compares to the energy needed for restructuring. When the impact energy is not enough to cause significant restructuring, particles grow by hit-and-stick collisions, and very fractal aggregates can be formed \citep{kempf1999}. When the impact energy is much larger, significant restructuring can take place, reducing the internal density of the dust aggregates. In this work, we will make use of the model presented in \citet{suyama2012} and \citet{okuzumi2012}. Specifically, we use Equation (15) of \citet{okuzumi2012} to calculate the volume of the a newly-formed aggregate, as a function of the masses and volumes of particles $i$ and $j$, the impact velocity, and the rolling energy $E_{\mathrm{roll}}$; the energy needed to roll two monomers over an angle of $90^\circ$ \citep{dominiktielens1997}.

\citet{gundlach2011} measured the rolling force between ice particles with radii of ${\sim}1.5\mathrm{~\mu m}$ to be $1.8\times10^{-3}\mathrm{~dyn}$, implying a rolling energy of $1.8\times10^{-7}\mathrm{~erg}$. Assuming the rolling force is size-independent \citep{dominiktielens1995}, the rolling energy is then often extrapolated using $E_{\mathrm{roll}}\propto a_0$. Recently however, \citet{krijt2014} showed that the rolling force scales with the size of the area of the monomers that is in direct contact, resulting in $F_{\mathrm{roll}}\propto a_0^{2/3}$, and $E_{\mathrm{roll}}\propto a_0^{5/3}$, leading to significantly smaller rolling energies when extrapolating down to monomer radii well below a micrometer. In this work, we use the scaling law of \citeauthor{krijt2014}, resulting in a rolling energy of $4\times10^{-9}\mathrm{~erg}$ for 0.1-$\mathrm{\mu m}$ radius ice particles. Physically, a \new{lower} rolling energy means less energy is needed to start restructuring of an aggregate. As a result, a \new{lower} rolling energy will lead to compacter aggregates.

\subsubsection{Gas and self-gravity compaction}\label{sec:gasgrav}
Aggregates can also be compressed by the ram pressure of the gas, or their own gravity, if they become very porous or massive. For low internal densities, \citet{kataoka2013a} found that the external pressure a dust aggregate can just withstand equals
\begin{equation}\label{eq:P_c}
P_c = \frac{E_{\mathrm{roll}}}{a_0^3}\phi^3.
\end{equation}
This pressure can then be compared to the pressure arising form the surrounding gas and from self-gravity
\begin{equation}\label{eq:P_gg}
P_\mathrm{gas} = \frac{v_\mathrm{dg} m}{\pi a^2 t_s}, \\
P_\mathrm{grav} = \frac{Gm^2}{\pi a^4},
\end{equation}
with $G$ the gravitational constant, in order to see whether an aggregate will be compacted as a result of these non-collisional processes \citep{kataoka2013c}. In this work, we will take these effects into account in a self-consistent way, while calculating the collisional evolution of the dust distribution.

\section{Monte Carlo approach}\label{sec:MC}
Numerical techniques for studying coagulation can be divided in two categories\footnote{See \citet{drazkowska2014} for a comparison between the two methods in the breakthrough growth case.}: integro-differential methods \citep[e.g.][]{weidenschilling1980,dullemonddominik2005,birnstiel2010}, and Monte Carlo (MC) methods \citep{gillespie1975,ormel2007,zsom2008}. Tracing particle porosity as well as mass becomes computationally expensive in the integro-differential approach. A solution to this issue was presented by \citet{okuzumi2012}, who assumed the porosity distribution for a given mass bin was narrow, but could vary in time. Since we are interested in including erosive processes, this assumption is not expected to hold, and for this reason we opt for the Monte Carlo method.

The approach to calculate the collisional evolution is based on the "distribution method" as described in \citet{ormel2008}. In this section we briefly revisit the method, focussing on what is new in this work. 

Let $f(\vec{x})$ be the (time-dependent) particle distribution function, with $\vec{x}_i$ the unique parameters describing dust particle $i$, in our case mass and filling factor\footnote{All other quantities (stopping time, volume, size, ...) can be calculated from these two numbers.}. For every pair of particles $i$ and $j$, one can determine the collision rate as
\begin{equation}
C_{ij} = K_{ij} / \mathcal{S},
\end{equation}
with $\mathcal{S}$ the surface area of the column\footnote{The size of the column is set by the total mass in the simulation and the dust surface density at the column's location.}, and $K_{ij}$ the collision kernel, which in this case equals
\begin{equation}\label{eq:K_z}
K_{ij} = \frac{\sigma_{ij}}{2\pi h_{d,i} h_{d,j}} \int_{-\infty}^{\infty} v_{\mathrm{rel}}(z) \exp\left( \frac{-z^2}{2h^2_{d,ij}} \right) dz,
\end{equation}
where $h_{d,i}$ is given by Equation \ref{eq:h_d}, and $h_{d,ij} = (h_{d,i}^{-2} + h_{d,j}^{-2})^{-1/2}$ and $\sigma_{ij}=\pi(a_i+a_j)^2$ equals the collisional cross section \citep{okuzumi2012}. This rate equation takes into account that particles with different properties inhabit different vertical scale heights, and is correct as long as the coagulation timescale is longer than the vertical settling/diffusion timescale. In this work, we approximate the integral over $z$ by assuming the mid plane relative velocity is a good indication for $v_\mathrm{rel}$ throughout the column. This allows us to solve the integral analytically and write
\begin{equation}
K_{ij} \simeq \frac{ \sigma_{ij}  h_{d,ij} }{ \sqrt{2\pi} h_i h_j} v_\mathrm{rel}(z=0).
\end{equation}
For the purpose of this paper, this approximation is sufficiently accurate, since most of the growth is expected to take place near the mid plane. 

Then, we can define the total collision rate for particle $C_i = \sum_{j>i} C_{ij}$, and the total collision rate $C_{\mathrm{tot}} = \sum_i C_{i}$. With all these rates known, 3 random numbers are used to identify which particles collide, and the time $\Delta t$ after which this collision occurs. The colliding particles are then removed from $f$, and the collision product is added. As a result, all collision rates $C_i$ have to be adjusted, since the particle distribution $f$ has changed. This cycle is then repeated.

The simple method has two main drawbacks. First, the time needed for updating the rates in between collisions scales with $N^2$, where $N$ is the total number of particles. Second, this method describes 1 collision per cycle, which can become a problem whenever the mass distribution is broad.

\subsection{Grouping method}
Rather than following every particle individually, identical particles can be grouped together. In our approach, the dust distribution is described by $N_f$ particle families. Within a single family, all particles have identical properties, in our case mass and internal density. In every family $i$, there are $w_i$ particle groups, each containing $2^{z_i}$ individual particles, where we call $z_i$ the zoom factor. The total number of particles in a single family therefor equals $g_i=w_i2^{z_i}$, and the total number of particles is $N=\sum_i g_i$. Instead of 2 particles colliding per cycle, collisions now happen between \emph{groups} of particles \citep[see][for details about this method]{ormel2008}.
Letting $i$ refer to the group with the lower zoom factor, we obtain for the group collision rates

\begin{equation}\label{eq:lambda_ij}
\lambda_{ij} = \begin{cases} 
w_i w_j 2^{z_i} C_{ij}  &\textrm{~~for~~} i \neq j,\\

w_i (w_i 2^{z_i}-1)C_{ii}  &\textrm{~~for~~} i = j,
\end{cases}
\end{equation}
where the $i=j$ case in Equation \ref{eq:lambda_ij} describe so-called \emph{in-group} collisions. Like before, we can define the total collision rate per family $\lambda_i = \sum_{j\geq i} \lambda_{ij}$, and the total collision rate $\lambda_{\mathrm{tot}}=\sum_i \lambda_i$, which can be used to determine which groups collide and when. This grouped approach has tremendous advantages, but there are also pitfalls, which we discuss in the following section.

\subsection{Sequential collisions}\label{sec:sequential}
Imagine the collision between a group of large bodies $i$ with a group of much smaller bodies $j$, such that $m_i \gg m_j$. Thus, a total of $2^{z_i}$ $i$-particles will collide with $2^{z_j}$ $j$-particles. Assuming $z_j \gg z_i$, every $i$-particle in the group will collide with $2^{z_j - z_i}$ $j$-particles in a single sequence before the collision rates are updated and the next groups to collide are chosen. We are assuming that the collision rates and the relative velocity between $i$ and $j$ particles are constant during this sequence. But this is only true if the properties of particle $i$ do not change significantly. For this reason we include the group splitting factor $N_\varepsilon$, that limits the number of collisions to $2^{z_j - z_i - N_\varepsilon}$.

Let $\delta m_i$ be the change in the mass of the larger particle $i$, after a single collision with a $j$-particle. Assuming the changes are small, we can then extrapolate to find the total change after the full sequence of collisions
\begin{equation}
\frac{\Delta m_i}{m_i} =  \frac{2^{z_j - z_i - N_\varepsilon} \delta m_i}{m_i}.
\end{equation}
Now, by imposing that $(\Delta m_i/m_i) \leq f_m$, we obtain
\begin{equation}\label{eq:N_e_m}
N^{(m)}_{\varepsilon} = \left[ -\log_2\left( \frac{f_m m_i 2^{z_i}}{\delta m_i 2^{z_j}}   \right)  \right],
\end{equation}
where the square brackets indicate that $N^{(m)}_\varepsilon$ is truncated to integers ${\geq}0$, which has the effect of particles with mass ratios ${\geq} f_m$ always colliding 1-on-1. In the case of perfect sticking, obviously $\delta m_i=m_j$, and Equation \ref{eq:N_e_m} reduces to Equation 12 of \citet{ormel2008}. We write an equivalent expression for the filling factor of the bigger grain 
\begin{equation}\label{eq:N_e_V}
N^{(\phi)}_{\varepsilon} = \left[ -\log_2\left( \frac{f_\phi \phi_i 2^{z_i}}{\delta \phi_i 2^{z_j}}   \right)  \right],
\end{equation}
where $\delta \phi_i$ denotes the change in $\phi$ after a single collision. The two limits are combined by writing
\begin{equation}
N_\varepsilon = \max\left( N^{(m)}_{\varepsilon} , N^{(\phi)}_{\varepsilon} \right),
\end{equation}
and ensure that neither the filling factor, nor the mass of the larger particle change by too much during a single Monte Carlo cycle. We note that $N_\varepsilon$ is not only a function of the masses and densities of both particles, but also of the relative velocity, since this influences $\delta \phi_i$ (and $\delta m_i$, when erosion is present). In this work, we will typically use $f_m = f_\phi =0.1$. 

Imposing this limit has two consequences. First, since the group of $i$-particles can now only collide with \emph{part of} the group of $j$-particles, this needs to be taken into account when the group collision rates are calculated, changing Equation \ref{eq:lambda_ij} into
\begin{equation}\label{eq:lambda_ij2}
\lambda_{ij} = \begin{cases} 
w_i w_j 2^{z_i+N_\varepsilon} C_{ij}  &\textrm{~~for~~} i \neq j,\\

w_i (w_i 2^{z_i}-1)C_{ii}  &\textrm{~~for~~} i = j.
\end{cases}
\end{equation}
Second, since it can occur that only part of a group collides, group numbers $w_i$ can now become fractional. This is fine as long as $w_i \geq 1$, ensuring that at least one full group collision can occur in the future \citep{ormel2008}.

\subsection{The distribution method}\label{sec:distribution_method}
For a given number of family members $g_i$, we have some freedom in choosing $z_i$; either creating many groups with few members (low $z_i$) or a few groups with many members (high $z_i$). This choice for the zoom-factors is crucial because it determines how many groups of a certain mass exist, which is related to the numerical resolution in that part of the mass range. Two approaches for determining the zoom-factors have been proposed by \citet{ormel2008}. 

One approach is the so-called "equal mass method", in which one strives to have groups of equal total mass. This method is essentially identical to the method of \citet{zsom2008}. With this approach, the peak of the mass distribution is very well traced, but parts of the particle distribution that carry little mass are described by few groups, resulting in larger uncertainties. The second option is the "distribution method", where one strives to have an equal number of groups per mass decade, independent of the total mass present in that interval. The difference between the two methods is nicely illustrated in Figure 4 of \citet{ormel2008}. Since we are interested in erosion, it is crucial to resolve the particle distribution over the entire mass range. It is for that reason that \new{we adopt the} distribution method.

In practice, this means that at certain times during the simulation, we calculate the total number of particles $N_{10}$ in every mass decade. The optimal zoom number for families in that mass range then equals
\begin{equation}
z^* = \left[ \log_2 \left( \frac{N_{10}}{w^*} \right) \right],
\end{equation}
where $w^*$ is the desired number of groups per mass decade. In this way, we construct a function $z^*(m)$, which gives the desired zoom number for a family with particle mass $m$. We then check every existing family: if a certain zoom number is too big, we "magnify" the group ($z_i \rightarrow z_i-1$, $w_i \rightarrow 2w_i$) until $z_i=z^*(m_i)$. Similarly, if the zoom number is too small, we "demagnify" ($z_i \rightarrow z_i+1$, $w_i \rightarrow w_i/2$). The (de)magnification process conserves particle number, but does force one to update the various collision rates. A more detailed description of (de)magnification is given by \citet{ormel2008}. In the rest of this work, we calculate and update the zoom factors after every $10^2$ collision cycles, whenever the peak or average mass has changed by ${>}5\%$, or when the maximum mass has changed by ${>}50\%$, which we found to ensure a smooth evolution of the zoom factors. We will use $w^*=60$ for the perfect sticking calculations, and $w^*=40$ for the ones including erosion.

\subsection{Merging}
Lastly, we have to address the merging of families. It can occur that demagnification results in a group number $w_i<1$, which is not allowed. When this occurs, the family does not contain enough individual particles to adopt $z_i=z^*(m_i)$. At this point, the family is insignificant. As we are simulating a fixed volume and the total mass needs to be conserved, we "merge" the family with another, 'healthy' (meaning $w_i>1$) one. First, we find the family $j$ that resembles family $i$ the most. In order to do so, we find the family that gives the largest product $(R^{(m)})(R^{(\phi)})^3$, where $R^{(\phi)} \leq 1$ is the ratio of the filling factors\footnote{This combination of $R^{(m)}$ and $R^{(\phi)}$ is used because the spread in masses is typically larger than the one in porosities, and we want to avoid merging particles with very different porosities if possible.}. Then, we merge the families into a new family $k$ with properties
\begin{equation}
g_k = g_i+g_j,\\
m_k = \frac{m_i g_i + m_j g_j}{g_i+g_j},\\
\phi_k = \frac{\phi_i g_i + \phi_j g_j}{g_i+g_j}.
\end{equation}
The new zoom- and group numbers are chosen such that $z_k=z^*(m_k)$. Merging is necessary to suppress the total number of groups.

\subsection{Non-collisional compaction}\label{sec:gasgrav_method}
Non-collisional compaction is implemented as follows: whenever a new aggregate is created in a collision, we calculate its compressive strength using Equation \ref{eq:P_c}, and compare this to the external pressures from gas ram pressure and self-gravity, calculated with Equation \ref{eq:P_gg} \citep{kataoka2013c}. If either one of the external pressures exceeds $P_c$, we compactify the dust grain (i.e. increase $\phi$) until the aggregate can withstand the external pressures.

\subsection{Erosion}\label{sec:erosion_method}
For every collision, we check if the conditions for erosion are met (i.e. $v_{\mathrm{rel}}>v_\mathrm{eros}$ and $R^{(m)}<R^{(m)}_\mathrm{crit}$), and if so, we determine the erosion efficiency using Equation \ref{eq:e_eros}. After a single erosive event, the mass that does \emph{not} end up in the target body equals $(1+\epsilon_\mathrm{eros})m_\mathrm{proj}$, see Figure \ref{fig:schem}. To limit the number of new families, we redistribute this mass over fragments with a mass of $m_\mathrm{frag} = m_\mathrm{proj}/10$. 

\section{Results}\label{sec:results}
In this section we show the results of our simulations for different erosion recipes, compaction mechanisms, turbulence strengths, and disk locations. When discussing the particle distribution at a given time, we shall use a number of quantities. These are the \emph{average} mass and porosity
\begin{equation}
m_a = \langle m_i \rangle, \\
\phi_a = \langle \phi_i \rangle,
\end{equation}
which trace the properties of the average particle, and the \emph{peak} mass and filling factor
\begin{equation}
m_p = \frac{ \langle m_i^2 \rangle }{ \langle m_i \rangle},\\
\phi_p = \frac{ \langle m_i\phi_i \rangle }{ \langle m_i \rangle},
\end{equation}
which trace the properties of the mass-dominating particle. We will also use the maximum mass $m_\mathrm{max}$, which is simply the mass of most massive particle.


\subsection{Perfect sticking}\label{sec:perfect_sticking}

\subsubsection{Collisional compaction only}\label{sec:perfect_sticking_a}
As a test for the Monte Carlo approach, we attempt first to match the trends observed in \citet{okuzumi2012}, who assumed perfect sticking between the dust grains. We adopt a turbulence strength parameter of $\alpha=10^{-3}$, and focus on a vertical column at 5 AU in a typical MMSN disk. At this point, \new{we} only include collisional compaction and omit erosion. To allow for a direct comparison to the work of \citeauthor{okuzumi2012}, we do not include the effects of Newton drag for particles with large Reynolds numbers in this simulation. In the rest of this work, Newton drag is always included self-consistently.

\begin{figure}[h!]
\centering
\begin{minipage}[b]{.48\textwidth}
\includegraphics[clip=,width=1.\linewidth]{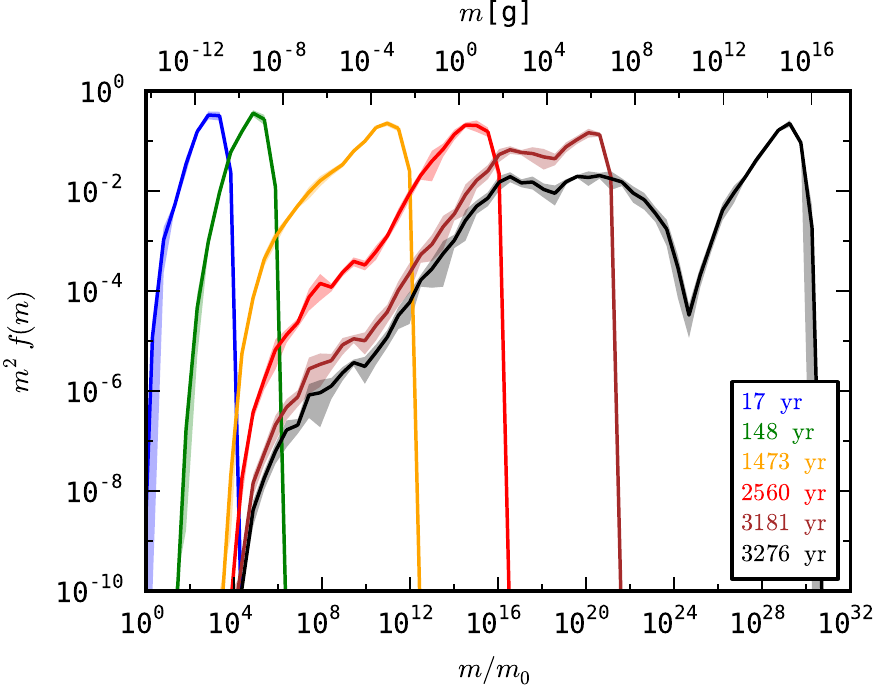}
\caption{Evolution of the normalized particle mass distribution at 5 AU with $\alpha=10^{-3}$, assuming perfect sticking and without compaction through gas and self-gravity. Only Epstein are Stokes drag are considered. Solid lines indicate averages over 4 Monte Carlo runs with identical starting conditions, and the shaded areas represent a spread of $1\sigma$.}
\label{fig:m2fm_sticking_5}
\end{minipage}\qquad
\begin{minipage}[b]{.48\textwidth}
\includegraphics[clip=,width=1.\linewidth]{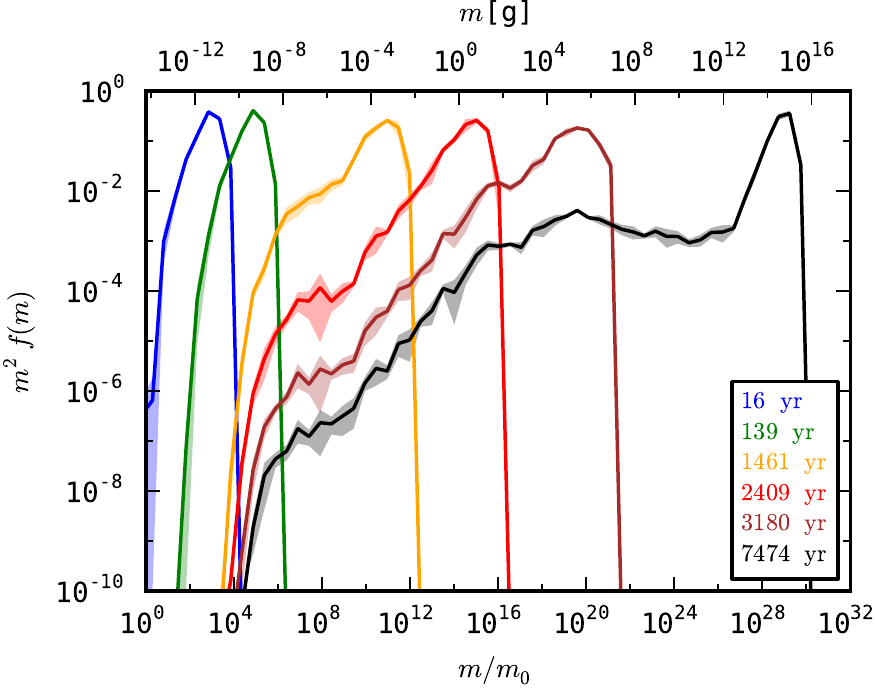}
\caption{Like Figure \ref{fig:m2fm_sticking_5}, but with compaction through gas and self-gravity and Newton drag for particles with $\mathrm{Re_p}>1$.}
\label{fig:m2fm_sticking_5_GG}
\end{minipage}
\end{figure}

\begin{figure}[h!]
\centering
\begin{minipage}[b]{.48\textwidth}
\includegraphics[clip=,width=1.\linewidth]{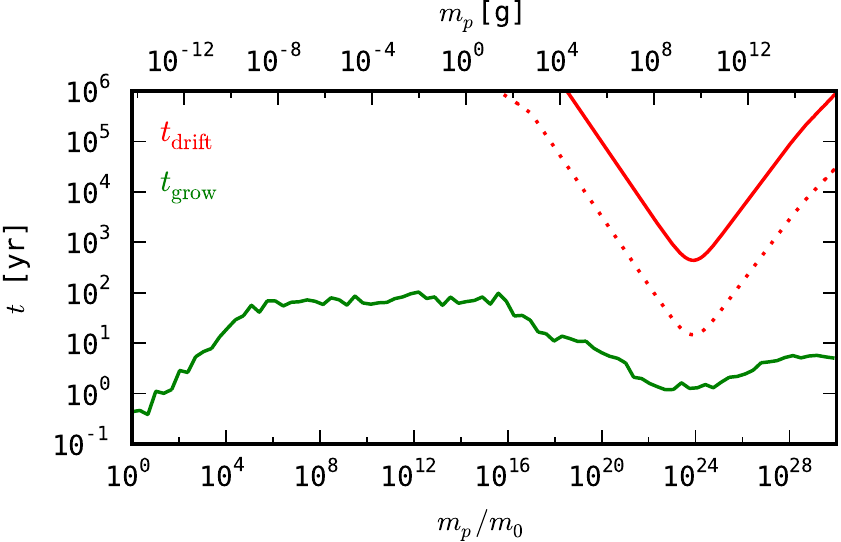}
\caption{Evolution of the growth- and radial drift timescale of the peak mass for the perfect sticking model at 5 AU with $\alpha=10^{-3}$. The dotted line indicates $(t_{\mathrm{drift}}/30)$. Only collisional compaction has been taken into account.}
\label{fig:drift_sticking_5}
\end{minipage}\qquad
\begin{minipage}[b]{.48\textwidth}
\includegraphics[clip=,width=1.\linewidth]{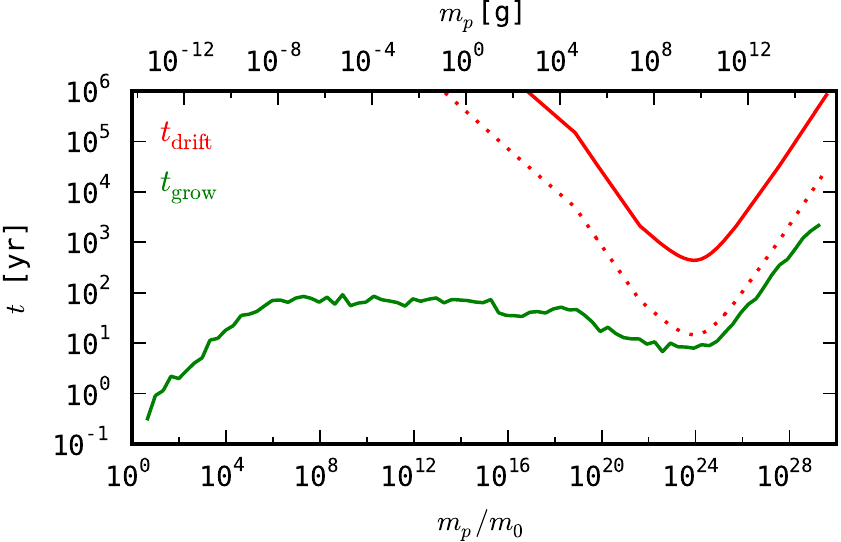}
\caption{Evolution of the growth- and radial drift timescale of the peak mass for the perfect sticking model at 5 AU with $\alpha=10^{-3}$. The dotted line indicates $(t_{\mathrm{drift}}/30)$. Compaction from gas and self-gravity, and Newton drag have been taken into account.}
\label{fig:drift_sticking_5_GG}
\end{minipage}
\end{figure}

Figure \ref{fig:m2fm_sticking_5} shows the evolution of the normalized mass distribution $m^2f(m)$ as a function of time. Solid lines mark the average over 4 Monte Carlo runs. Thanks to the distribution method described in Section \ref{sec:distribution_method}, the sampling of the mass distribution is very good over the entire mass range: even at later times, when most of the mass is located in particles with masses of ${\sim}10^{15}$ g, the distribution of particles all the way down to $10^{-9}$ g is resolved remarkably well, despite these particles only making up a very small fraction of the total mass.

When we compare Figure \ref{fig:m2fm_sticking_5} to Figure 7 of \citet{okuzumi2012}, it is clear that our local MC method yields very similar results. We recognize the familiar narrow mass peak when growth is governed by Brownian motion, followed by a broader distribution once turbulence kicks in. Once particles reach $\Omega t_s\sim1$ ($m_j\simeq10^{10}\mathrm{~g}$ in this case), systematic drift greatly increases their collision rate, and very rapid growth ensues. The slight difference in timescales is attributed to $i)$ the slightly different value for the rolling energy, $ii)$ our approximation of Equation \ref{eq:K_z}, and $iii)$ our use Equation \ref{eq:t_s} to calculate the stopping times, while \citeauthor{okuzumi2012} used $t_s = t_s^\mathrm{(Ep)} + t_s^\mathrm{(St)}$ to ensure a smooth transition between Epstein and Stokes drag (S. Okuzumi, private communication).

While we take into account drift-induced relative velocities, the dust particles are bound to our simulated column and cannot move radially through the disk. To test the validity of this assumption, we compare the growth timescale of the peak mass, defined as
\begin{equation}
t_{\mathrm{grow}} \equiv \frac{m_p}{(d m_p /dt)},
\end{equation}
to the radial drift timescale at that mass
\begin{equation}
t_{\mathrm{drift}} \equiv \frac{R}{v_{\mathrm{drift}}(m_p)}.
\end{equation}
The radial drift velocity is given by \citep{weidenschilling1977}
\begin{equation}
v_\mathrm{drift} = - \frac{2\Omega t_s}{1+(\Omega t_s)^2} \eta v_K,
\end{equation}
where $v_K=R\Omega$ is the Keplerian orbital velocity, and $\eta$ can be written as \citep{nakagawa1986}
\begin{equation}\label{eq:eta}
\eta \equiv - \frac{1}{2} \left( \frac{c_s}{v_K} \right)^2 \frac{\partial \ln(\rho_g c_s^2)}{\partial \ln R} = 4\times10^{-3} \left(\frac{R}{5\mathrm{~AU}}\right)^{1/2}. 
\end{equation}

Figure \ref{fig:drift_sticking_5} shows both the growth and radial drift timescales during the complete evolution of the peak mass. Initially, relative velocities are dominated by Brownian motion. Since this velocity drops with increasing particle mass, the growth timescale increases. Around a mass of $10^{-9}$ g, turbulent velocities start to dominate the relative velocity, and the growth timescale stays approximately constant. Particles larger than $10^3$ g enter the second turbulent regime as $t_s(m_p)>t_\eta$. In this regime, velocities between similar particles are increased (see Equation \ref{eq:v_turb}), which leads to a decrease in the growth timescale. Since the growth timescale is always much smaller than the drift timescale, the aggregates in this simulation do indeed out-grow the radial drift barrier.

\subsubsection{Including gas and self-gravity compaction}\label{sec:perfect_sticking_b}
The next step is to include compaction by gas pressure and self-gravity, as described in Section \ref{sec:gasgrav}. In addition, we now take into account Newton drag for particles with large Reynolds numbers. Figure \ref{fig:m2fm_sticking_5_GG} shows the results for the same disk parameters as before. The general shape of the evolution looks similar to Figure \ref{fig:m2fm_sticking_5} initially, but from the corresponding times it is clear that the growth is slower for the largest aggregates. The main reason for this is that the largest dust grains are compacted by the gas and self-gravity, resulting in a smaller collisional cross section. In addition, the aerodynamic properties are different, which affects the relative velocities.

The growth- and drift timescales are plotted in Figure \ref{fig:drift_sticking_5_GG}. When we compare Figures \ref{fig:drift_sticking_5} and \ref{fig:drift_sticking_5_GG}, we confirm that the growth close to the drift barrier is slower when using the full compaction recipe. For the largest particles, the growth timescale is increased by more than 2 orders of magnitude. In addition, including Newton drag has broadened the drift barrier somewhat. Nonetheless, the growth is still fast enough to prevent particles from drifting significant distances. 

\subsubsection{Evolution of internal densities}
\begin{figure}
\includegraphics[clip=,width=1.\linewidth]{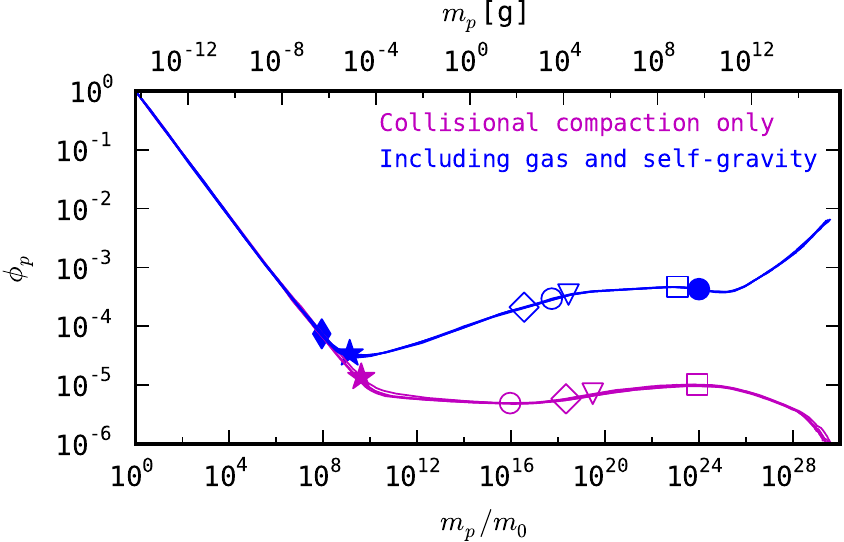}
\caption{Evolution of the internal structure of the mass-dominating particles, for the perfect sticking models at 5 AU, for the models with and without non-collisional compaction mechanisms. Aggregates start out as monomers in the top left corner, and grow towards larger sizes and porosities. Lines show individual simulations. Open symbols correspond to points where the mass dominating particles reach $a = \lambda_{\mathrm{mfp}}~~(\circ)$; $t_s = t_\eta~~(\Diamond)$; $\Omega t_s = \alpha~~(\triangledown)$; and $\Omega t_s = 1~~(\square)$. Filled symbols show peak mass and filling factor at the times of first: collisional compaction $(\star)$; gas-pressure compaction $(\blacklozenge)$; and self-gravity compaction $(\bullet)$.}
\label{fig:internal_5}
\end{figure}

It is interesting to compare the evolution of the internal densities of the particles for the models with and without non-collisional compaction. In Figure \ref{fig:internal_5}, the peak filling factor is plotted versus the peak mass for the simulations described so far. The symbols correspond to important points in the evolution of the aggregates: open circles are related to the stopping time of the aggregates, and closed symbols indicate the onset of various compaction mechanisms\footnote{Note that the particle actually undergoing this compaction can have a mass and porosity that differ slightly from $m_p$ and $\phi_p$.}.

Initially, aggregates grow through hit-and-stick collisions, and evolve along a line of constant fractal dimension close to 2. In the collisional-compaction-only scenario, particles reach a filling factor of ${\sim}10^{-5}$ during hit and stick growth, before collisional compaction kicks in, after which $\phi$ stays almost constant. When $\Omega t_s(m_p)>1$, the internal density drops even further. The general picture, as well as the location of the various turnover points, is consistent with the top panel of Figure 10 of \citet{okuzumi2012}. When non-collisional compaction is included, the filling factor, in general, is much higher at later times, and follows the boundaries that have been described by \citet{kataoka2013c} (e.g. their Figure 3). For this particular combination of turbulence, rolling energy, and monomer size, compacting by gas ram pressure actually occurs \emph{before} the first collisional compaction event takes place\footnote{In fact, the gas compaction starts when the aggregates are still in the Epstein drag regime. Equation \ref{eq:P_c} is determined by \emph{static} compression of porous aggregates, and Equation \ref{eq:P_gg} assumes the external pressure can be treated as continuous. However, if the collision frequency of gas molecules with individual monomers of the aggregate is low compared to the frequency at which monomer-monomer contacts oscillate and dissipate energy, this approach might not be accurate. Future work is encouraged to investigate the effect of collisions between the aggregate and gas molecules in this regime.}. Significant settling occurs when $\Omega t_s > \alpha$, which corresponds to $m\sim10^{-3}$ g for compact particles (see Figure \ref{fig:stokes}). From Figure \ref{fig:internal_5} however, we see that porous particles only begin to settle when their masses reach ${\sim}10^4-10^5$ g. Lastly, aggregates with masses above ${\sim}10^{10}\mathrm{~g}$ are compacted by self-gravity, causing the filling factor for the largest bodies to be several orders of magnitude higher. In the remainder of this work, we include both collisional and non-collisional compaction mechanisms, and Epstein, Stokes, and Newton drag self-consistently.

\subsection{Erosion}\label{sec:erosion_sims}
\begin{figure}
\includegraphics[clip=,width=1.\linewidth]{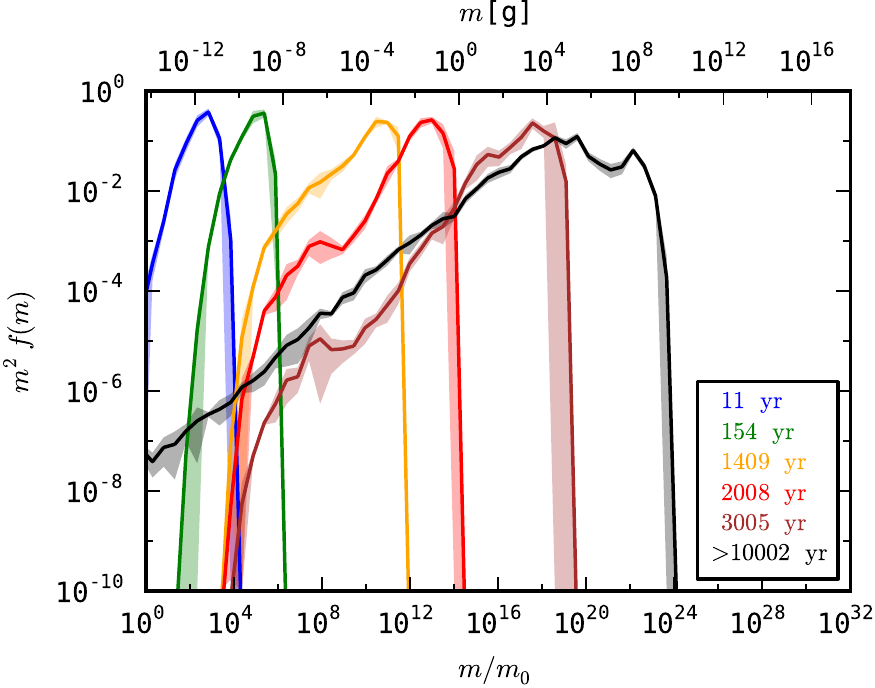}
\caption{Evolution of the normalized particle mass distribution at 5 AU with $\alpha=10^{-3}$, assuming $v_{\mathrm{eros}}=20\mathrm{~m~s^{-1}}$. The full compaction model is used.}
\label{fig:m2fm_E1}
\end{figure}

With this framework in place, the final step is to include the erosion model of Section \ref{sec:erosion_model} in the simulations, and calculate the evolution of the particle distribution self-consistently. Figure \ref{fig:m2fm_E1} shows the mass distribution at various times for $v_{\mathrm{eros}}=20\mathrm{~m~s^{-1}}$. Initially, the evolution proceeds just like in \ref{fig:m2fm_sticking_5_GG}, but as the largest aggregates approach $\Omega t_s=1$, their velocity relative to smaller particles is high enough for erosion, and their growth stalls. As a direct consequence of the erosion, the amount of small particles increases, and after ${\sim} 4000\mathrm{~yr}$ a steady-state is reached, with a significant amount of mass residing in particles smaller than a few grams.

\begin{figure}
\includegraphics[clip=,width=1.\linewidth]{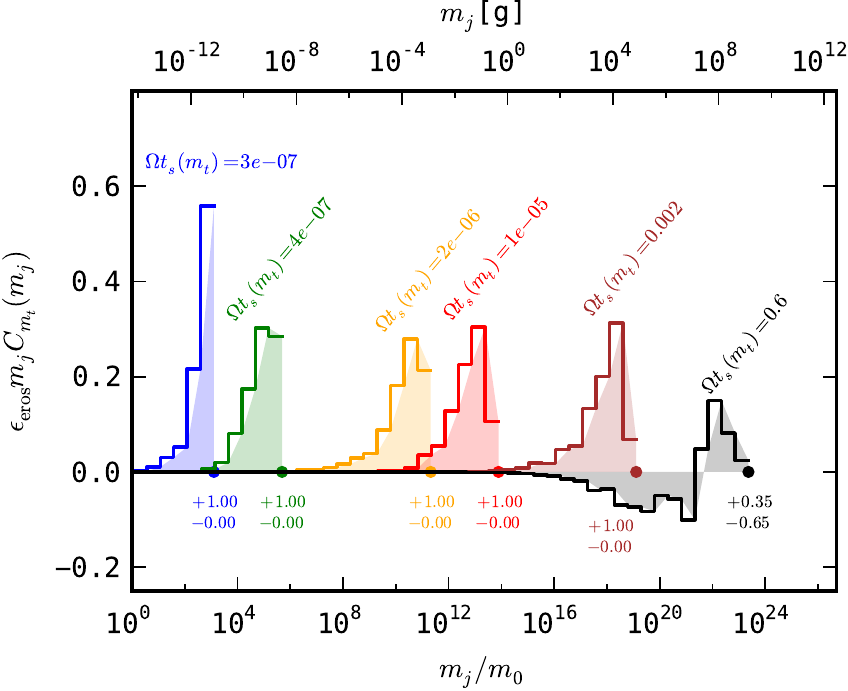}
\caption{Projectile distribution mass functions for simulation E1, constructed for the \emph{maximum} masses ($m_t=\bullet$) at various times. Colors and times correspond to Figure \ref{fig:m2fm_E1}. For each distribution, the stopping time of the $m_t$-particle is given, and the weights of the total positive and negative area are plotted. The distributions have been normalized in such a way, that the absolute sum of the contributions equals 1.}
\label{fig:projectile_E1}
\end{figure}

To investigate how erosion halts the growth of the largest bodies, it is instructive to plot so-called projectile mass distributions \citep{okuzumi2009}. For a certain particle mass $m_t$, these distributions show the contribution to the growth of that particle as a function of projectile mass $m \leq m_t$. An example of such a plot is shown in Figure 9 of \citet{okuzumi2012}, where the distribution function is plotted at various times for $m_t=m_p$. For our distribution plots, we make two important changes: First, since we are interested in the growth of the largest bodies, we plot projectile distributions for $m_t = m_{\mathrm{max}}$, with $m_{\mathrm{max}}$ the \emph{largest} mass in the simulation at a given time. Second, to illustrate the  effect of erosive collisions, we calculate the mass loss for every erosive collision, taking into account the correct erosion efficiency\footnote{A sticking collision, where the mass of the projectile is added to the target, is described by $\epsilon_\mathrm{eros}=-1$.}. As a result, the sign of the distribution function can be both positive and negative. Figure \ref{fig:projectile_E1} shows the distribution for one of the simulations of Figure \ref{fig:m2fm_E1} (colors correspond to the same times). When we examine the right-most projectile mass distribution, corresponding to a time $t=10^4\mathrm{~yr}$, it is immediately clear how erosion affects the evolution of particles with $\Omega t_s \sim 1$. While these aggregates grow by collisions with similar-sized bodies, they lose mass by colliding with particles that have a mass below $10^{-2}m_t$. This could have been predicted by looking at Figure \ref{fig:velocities}, from which it is clear that the highest velocities are attained between particles with mass ratios well below unity. The importance of this erosion however, depends on the current mass distribution, and can only be tested through dedicated simulations like the ones presented here. Since the area under the negative part of the projectile distribution outweighs the positive part, the erosion is so effective that it stops the growth of the largest bodies, resulting in the behavior seen in Figure \ref{fig:m2fm_E1}. 


We define a parameter $\zeta$ using the positive and negative areas under the projectile mass distributions
\begin{equation}
\zeta = \frac{ \sum C_+ - \sum C_- }{ \sum C_+ + \sum C_- }, 
\end{equation}
with $\sum C_{+}$ and $\sum C_-$ the sums of the positive and negative part of the projectile mass distribution respectively. The parameter $\zeta$ ranges from 1 (no erosion) to -1 (only erosion), and equals 0 when there is a balance between growth and erosion.

\begin{figure}
\includegraphics[clip=,width=1.\linewidth]{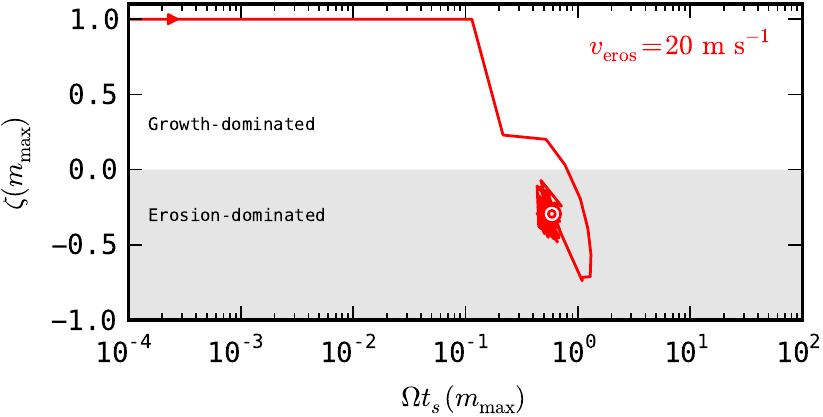}\\
\includegraphics[clip=,width=1.\linewidth]{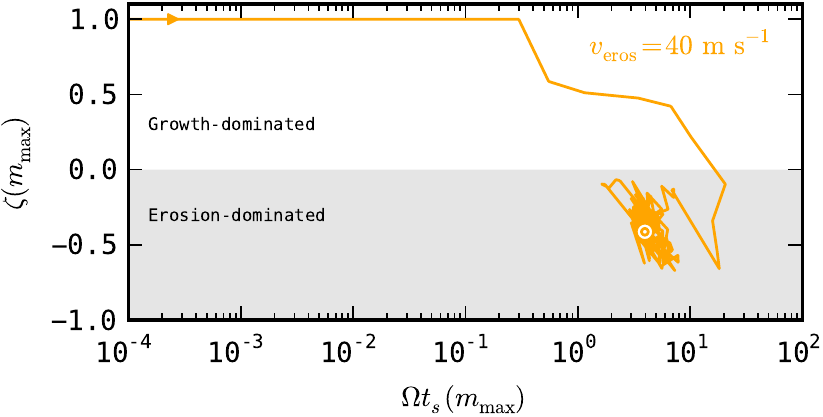}\\
\includegraphics[clip=,width=1.\linewidth]{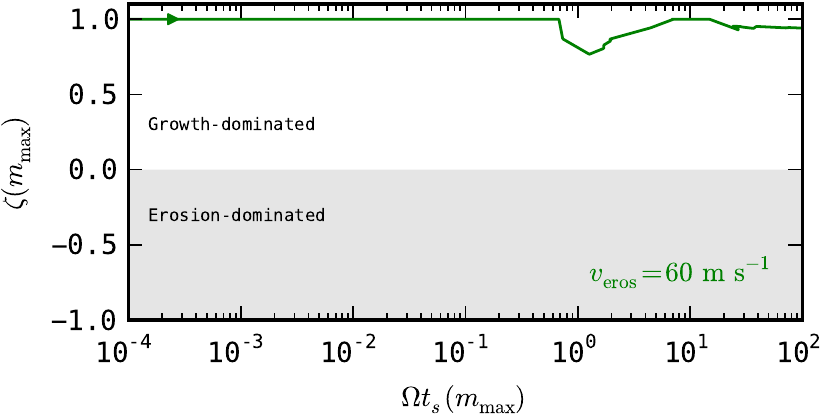}
\caption{Evolution of $\zeta(m_\mathrm{max})$ for erosive simulations with $v_\mathrm{eros}=20,40,60\mathrm{~m~s^{-1}}$ as a function of $\Omega t_s(m_\mathrm{max})$, showing the impact of erosion on the ability of the largest bodies to grow. In the upper two panels, the steady-state is indicated by the $\circledcirc$-symbol.}
\label{fig:zeta}
\end{figure}

The top panel of Figure \ref{fig:zeta} shows the evolution of $\zeta$ for the most massive particle during one of the simulations of Figure \ref{fig:m2fm_E1}, plotted as a function of $\Omega t_s$ of the maximum mass. Early on, there is no erosion present and $\zeta=1$, but as the largest bodies grow towards $\Omega t_s=1$, erosion increases and $\zeta$ drops. When $0<\zeta<1$, the massive particles still grow faster then they are eroded, but the erosion can be significant in that it results in the creation of more small particles, thus increasing its destructive effect. When $\zeta<0$, erosion dominates over growth and the most massive particles are loosing considerable mass. This causes the curve in Figure \ref{fig:zeta} to turn around. As bodies shrink, there is less erosion and $\zeta$ increases again. A quasi steady-state is reached with $\zeta$ just below unity and $\Omega t_s(m_\mathrm{max}) \sim 0.6$. The reason $\zeta \neq 0$ during the steady state, is that it is not the same particle that is the most massive at all times. Instead, particles take turn at being the most massive body. Since the largest particles are stuck at a mass and size for which drift is fastest, they will move radially towards the central star. For this combination of parameters, we conclude that growth beyond the drift barrier is impeded by erosion. 

The other panels of Figure \ref{fig:zeta} show similar plots but for different erosion threshold velocities. For $v_\mathrm{eros}=40\mathrm{~m~s^{-1}}$ (middle panel), erosion is less efficient and the largest bodies grow to $\Omega t_s\simeq10$ before they start to lose mass rapidly. The reason particles can grow larger is twofold. First, the threshold velocity itself is somewhat higher, causing erosion to start for higher masses. Second, since the erosion efficiency is proportional to $(v_\mathrm{rel}/v_\mathrm{eros})$, the high-velocity projectile are less efficient in excavating mass from the targets. Both effects together cause the largest mass in the steady state to be about a factor of 10 larger than in the top panel of Figure \ref{fig:zeta}. Finally, the bottom panel shows the results for $v_\mathrm{eros}=60\mathrm{~m~s^{-1}}$. This is a special case, since now the erosion threshold velocity can only be reached around $\Omega t_s=1$, with radial drift, azimuthal drift, and turbulence contributing (see Figure \ref{fig:velocities}). Indeed, erosion is strongest around $\Omega t_s=1$, but is inefficient and $\zeta$ never drops below 0. When $\Omega t_s>20$, erosion reappears, as a result of smaller particles drifting into the larger bodies. But, since $\zeta \sim 1$, bodies can continue to grow relatively unaffected.


\begin{figure*}
\centering
\includegraphics[clip=,width=.95\linewidth]{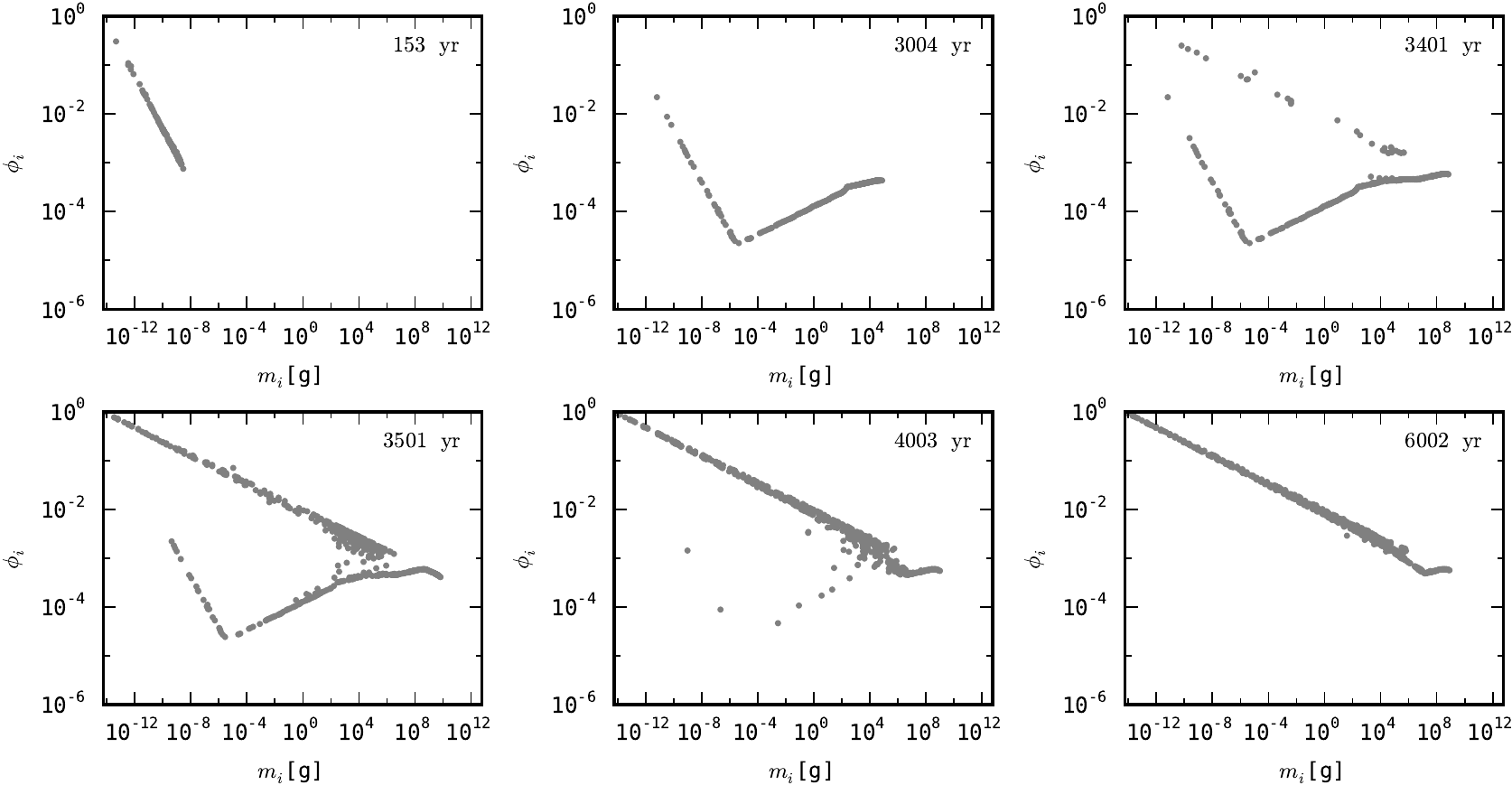}
\caption{Masses and filling factors of all unique families at different times, for simulation E1. One dot corresponds to one family, and does not provide information about the total mass or number of members in that family.}
\label{fig:fams}
\end{figure*}

\subsubsection{Variation in porosity}\label{sec:varpor}
One of the biggest advantages of the Monte Carlo method is that aggregate mass and porosity are treated truly independently. In other words, aggregates of identical mass can have a very different porosity. However, the collision model used in this work immediately implies that the spread in porosities (for a given particle mass) will be narrow, when sticking collisions dominate the evolution. For example, the collision model, at the moment, does not include an impact-parameter dependence in collisions, or a random component in the relative velocity. As a result, collisions between particles with certain properties always occur at the same relative velocity, and always result in the same collision product(s). Moreover, when gas compaction (or self-gravity compaction) limits the porosity of an aggregate, bodies will evolve along $P_c=P_\mathrm{gas}$ (or $P_c=P_\mathrm{grav}$), according to Equations \ref{eq:P_c} and \ref{eq:P_gg}. As a result, mass-porosity relations as shown in Figure \ref{fig:internal_5} accurately represent the internal structure of the majority of aggregates. 

This picture changes when erosion starts to play a role. Figure \ref{fig:fams} shows the evolution of the properties of each family in one of the E1 simulations. Note that each dot corresponds to a single family, and that the total masses and number of family members can vary significantly between families. Nonetheless, Figure \ref{fig:fams} gives a good indication of the spread in porosity. For the reasons described above, the spread in porosity is very small during the first 3000 years of the evolution. After 3400 years, the first erosive collisions have occurred, and created a population of fragments with a fractal dimension set by the parent body. At this point, the porosity distribution becomes bimodal, and the assumption of a single porosity parameter - which only depends on aggregate mass - is untenable. Later, after ${\sim}6000$ years, the original population of aggregates, whose porosity was set by their growth history, has disappeared. A steady-state is reached in which the internal structure of the fragments is dominated by the porosity of the particles that act as targets for erosion, i.e. the large bodies with $\Omega t_s \sim 1$.

\section{Semi-analytical model}\label{sec:semian}
The evolution of the mass-dominating particles can be captured in a simple semi-analytical model. Assuming the entire dust mass is located in particles of identical mass $m_p$, the growth rate can be written as \citep{okuzumi2012}
\begin{equation}\label{eq:dmdt}
\frac{dm_p}{dt} = \frac{\Sigma_d}{\sqrt{2\pi}h_d}\sigma_{\rm col} v_{\rm rel}.
\end{equation}
The collisional cross section depends directly on the particle porosity, and the relative velocity and dust scale height depend on $\phi$ through the particle stopping time. As a simple model for the aggregate's internal structure, we assume the aggregates initially grow with a constant fractal dimension of ${\sim}2$, until the kinetic energy in same-sized collisions exceeds $E_\mathrm{roll}$. After that, the internal structure can be calculated through Equation 31 of \citet{okuzumi2012}, but in practice is always dominated by the gas/self-gravity compression of \citet{kataoka2013c}, see Section \ref{sec:gasgrav}.

This approach, \new{similar to \citet[][Section 5.3]{kataoka2014}}, is valid when particles grow primarily  through collisions with similar-sized particles. This is valid in most regimes, but not true in the first turbulence regime. Here, relative velocities between identical particles are suppressed, and aggregates grow by collecting smaller particles. However, it can be shown that in this regime the growth timescale is approximately constant \citep{okuzumi2009}. Hence, we will assume that $t_\mathrm{grow}$ is constant in the regime where turbulence dominates $v_\mathrm{rel}$, and $t_s<t_\eta$.

\begin{figure*}
\centering
\includegraphics[clip=,width=.95\linewidth]{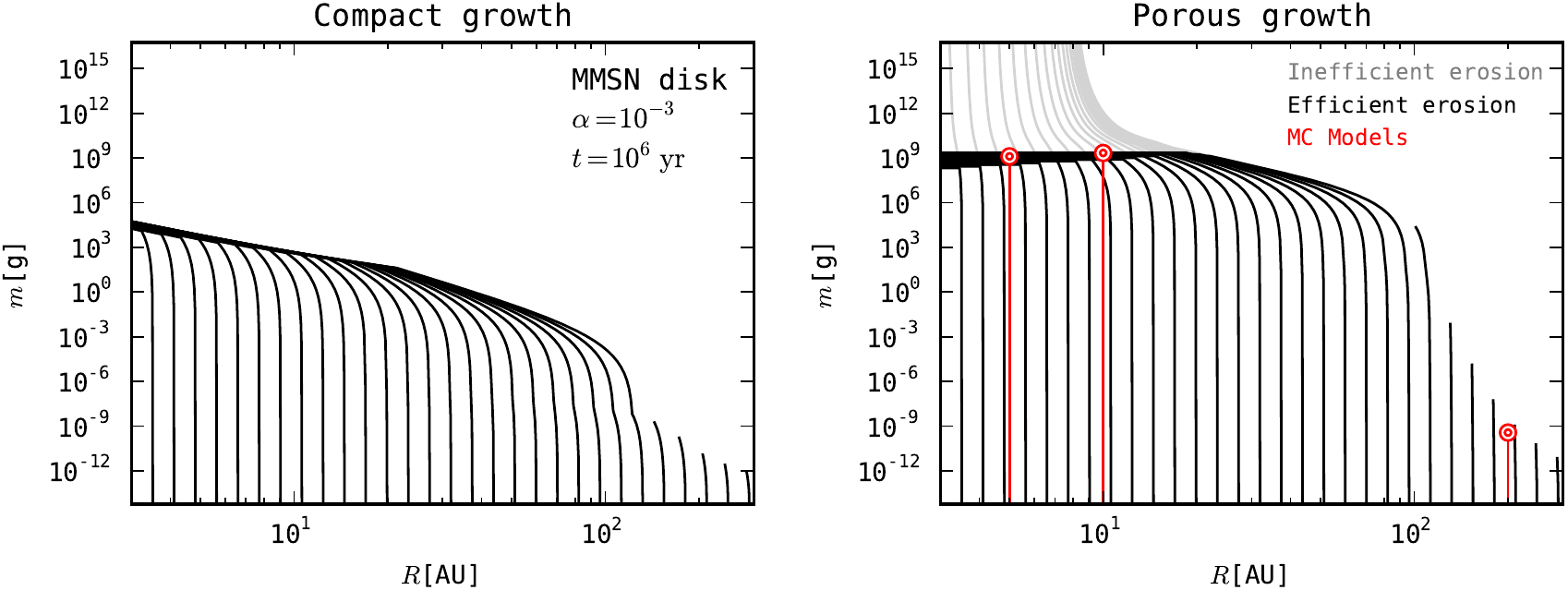}
\caption{Evolution of $m_p(t)$ and $R(t)$ for dust coagulation as obtained from the semi-analytical model (Equations \ref{eq:dmdt} and \ref{eq:dRdt}), for an MMSN disk and $\alpha=10^{-3}$. Lines indicate different starting conditions $R(t=0)$, and are evolved for $10^6$ yrs. \emph{Left:} Compact growth: $\phi=1$ at all times, and $v_\mathrm{frag}=10\mathrm{~m~s^{-1}}$. \emph{Right:} Porous growth: the internal structure of the aggregates is set by hit and stick growth, followed by collisional compaction or gas and self-gravity compaction. Grey lines have no erosion, while black lines show the results for $v_\mathrm{eros}=40\mathrm{~m~s^{-1}}$. Colored lines and $\circledcirc$-symbols indicate the evolution and steady state peak mass obtained through local Monte Carlo simulations (Section \ref{sec:erosion_sims}).}
\label{fig:semian2D}
\end{figure*}

At the same time, the radial drift of the particles is governed by 
\begin{equation}\label{eq:dRdt}
\frac{dR}{dt} = -v_\mathrm{drift},
\end{equation}
with the drift velocity a function of $\Omega t_s$. Assuming a fixed dust to gas ratio of $10^{-2}$ throughout the disk, we can solve Equations \ref{eq:dmdt} and \ref{eq:dRdt} to obtain the evolution of the vertically integrated peak mass. Catastrophic fragmentation is taken into account by setting $(dm_p/dt)=0$ when $v_\mathrm{turb}>v_\mathrm{frag}$ for two particles of mass $m_p$. Figure \ref{fig:semian2D} shows lines along which the dust evolves, starting from $m=m_0$ at various locations in the disk. The left plot shows the results for compact growth (i.e. $\phi=1$ at all times), after $10^6$ yr. (For the compact case, we have temporarily set $v_\mathrm{frag}=10\mathrm{~m~s^{-1}}$.) Initially, growing aggregates are not moving radially, resulting in vertical lines in Figure \ref{fig:semian2D}. As particles' Stokes numbers increase, collision velocities and drift speeds increase. In the inner regions of the disk, the maximum size is limited by fragmentation through same-sized collisions. Particles cannot grow larger than ${\sim}$cm, and will inevitably drift inwards. In the intermediate region, from $20-100$ AU, the fragmentation velocity is not reached. Here, the maximum size is set by radial drift. In the outermost disk (beyond $10^2$ AU), growth is very slow because of the low dust densities, and $10^6$ yr is not enough to reach the size necessary to start drifting. The general behavior is identical to what is observed in full compact coagulation models (cf. Figure 3 of \citet{testi2014}).

The grey lines in the right-hand panel of Figure \ref{fig:semian2D} show the results of the semi-analytical model for porous growth, where $\phi$ is set by collisional, gas pressure, and self-gravity compaction, assuming perfect sticking. Since we are assuming the $m_p$ particles carry the total dust mass, we do not have any information about the mass-distribution of smaller particles. Nonetheless, we can mimic the effect of effective erosion, by setting $(dm_p/dt)=0$ when the relative velocity between the mass dominating particle and small projectiles (taken to be monomers) exceeds $v_\mathrm{eros}$. The black solid lines in the right panel of Figure \ref{fig:semian2D} show the results for $v_\mathrm{eros}=40\mathrm{~m~s^{-1}}$, while the red lines indicate results for the peak mass of the full Monte Carlo models for the same erosion threshold velocity (note that the maximum mass reached in these models can be a factor of ${\sim}10$ larger). We have also included a full model run at 200 AU, which we evolved for $10^6$ yrs. The results of the semi-analytical model agree with the simulations of the previous section remarkably well. 

\section{Discussion}\label{sec:discussion}
From the maximum sizes fluffy aggregates can reach at a given location, we identify three zones in the protoplanetary disk:
\begin{itemize}
\item{\emph{3-10 AU:} Assuming perfect sticking, the combination of Stokes drag and enhanced collisional cross sections allows the porous aggregates in the inner disk to out-grow the radial drift barrier, and reach planetesimal sizes without experiencing significant drift. However, when erosion is efficient, mass loss in erosive collisions stalls the growth around $\Omega t_s\sim1$, preventing the porous aggregates from crossing the radial drift barrier (Figure \ref{fig:zeta}).}
\item{\emph{10-100 AU:} At intermediate radii growth timescales increase and radial drift takes over, even before aggregates reach sizes and stopping times that allow for erosive collisions to take place.}
\item{\emph{>100 AU:} In the outer disk, the disk lifetime is not long enough for particles to grow to sizes where significant drift occurs. In the porous growth scenario, aggregates this far out are in the hit-and-stick regime, and their surface-to-mass ratio does not change when they gain mass. As a result, hardly any drift is visible. In the compact case, an increase in mass automatically results in a decrease in the surface-to-mass ratio, and the onset of radial drift is already visible for very low particle masses.}
\end{itemize}

For erosion to start, the collision velocity between target and projectile needs to exceed $v_\mathrm{eros}$. In the limit where the projectiles are monomers that couple to the gas extremely well, this collision velocity equals the relative velocity of the large bodies with respect to the gas. When the largest particle has $\Omega t_s \gg 1$, it moves on a Keplerian orbit, and $v_\mathrm{dg}\simeq \eta v_K$, while bodies with $\Omega t_s =1$ have a slightly larger velocity with respect to the gas \citep{weidenschilling1977}. For the disk model employed in this work (Section \ref{sec:disk}), the quantity $\eta v_K$ does not depend on $R$, and thus the maximum drift speed is constant though out the disk. It is clear then from Equation \ref{eq:eta} that growing aggregates in colder disks (lower $c_s$), or disks with a (locally) shallower gas density profile might suffer less from erosion.

\new{It is clear from Figure \ref{fig:zeta} that the size of $v_\mathrm{eros}$ is a key parameter: its value, together with $\eta v_K$, determines whether growth beyond $\Omega t_s = 1$ is possible or not. Unfortunately, the value of $v_\mathrm{eros}$, or even its relation to $v_\mathrm{frag}$, is not accurately known for the large and highly-porous icy bodies in question (Section \ref{sec:erosion_model}). Numerical investigations, showing conflicting trends for erosion efficiency with mass ratio, often employ monodisperse grain sizes \citep[e.g.,][]{seizinger2013c,wada2013}, and the threshold velocities depend almost linearly on the grain radius (Equation \ref{eq:v_frag}), a parameter which itself is not well constrained. At the same time, the only available experimental work on erosion for ices used a distribution of grain sizes \citep{gundlach2014}. In addition, both numerical and experimental studies are restricted to sizes ${\lesssim}\mathrm{mm}$ and porosities ${\gtrsim}10^{-1}$, and cover a soberingly small portion of the parameter space encountered in this work (e.g., Figure \ref{fig:internal_5}). Future studies, numerical as well as experimental, are encouraged to elucidate these matters, and constrain the threshold for erosion and its dependence on target/projectile sizes and porosity. Finally, we assume that material that is eroded locally is removed from the target. In reality, the fate of the fragments will be determined by the local gas flow and the velocity with which they are ejected. For very porous targets, the gas flow through and around the surface of the target might result in these fragments being re-accreted \citep{wurm2004}. If efficient, this re-accretion might be a way to alleviate the destructive influence of erosive collisions. On the other hand, the flow through a body is likely to be insignificant, unless it is extremely porous \citep{sekiya2005}.}

So far we have assumed that while erosion can play an important role, catastrophic fragmentation does not occur. The maximum velocity between same-sized bodies is reached for $\Omega t_s=1$, and equals ${\sim} (3/2) \alpha^{1/2}c_s$ (Equation \ref{eq:v_turb}). Since the sound speed diminishes for increasing radii, this velocity is highest in the inner disk. For typical turbulence strengths ($\alpha \lesssim 10^{-3}$) and small icy monomers, this velocity will not exceed the fragmentation threshold velocity (Equation \ref{eq:v_frag}), \new{and, especially in the outer disk, fragmentation of icy bodies through catastrophic fragmentation is very unlikely. However, if \emph{all} collisions result in sticking, small particles (${\lesssim}100\mathrm{~\mu m}$) are removed from the protoplanetary nebula very rapidly, contradicting observational constraints \citep{dullemonddominik2005,dominikdullemond2008}. Drift-induced erosion can alleviate these issues, since the maximum drift velocity is high throughout the entire disk.}

In this work, we have assumed collisions below the fragmentation threshold to result in perfect sticking, \new{i.e. the mass of the resulting aggregate equals the sum of both colliding masses.} However, even for collisions below the fragmentation threshold velocity, a significant amount of mass may be ejected during a collision, especially if the collision occurs at a large impact parameter \citep{paszun2009,wada2013}. An advantage of a Monte Carlo model approach like the one presented here, is that it is relatively straightforward to include an additional random number to determine, for example, the impact parameter. The difficulty lies in obtaining a collision model that describes the collisional outcome as a function of said parameter. A good start would be the work of \citet{wada2013}, who show the growth efficiency as a function of impact parameter (Figure 4). Basically, head-on collisions promote growth, while collisions with a large impact parameter result in little mass gain. Unfortunately, much less is known about the porosities of the resulting aggregates.

\new{At the heart of the model of Section \ref{sec:MC} lies the assumption that an aggregate is adequately described by two quantities: its mass and (average) porosity. While this represents a considerable improvement on the compact coagulation assumption, a single average porosity does not allow for a complex internal structure of the aggregates. For small grains, the accuracy of this assumption will depend on their collisional history. For example, one can imagine a porous aggregate with a denser outer shell being formed if the aggregate is compacted through many collisions with small mass ratios \citep{meisner2012}. Such a compact rim will hardly alter the aggregate's average porosity, but can influence its sticking and erosion behavior \citep{schrapler2011}. Likewise, gas- and self-gravity compaction need not result in a homogenous internal structure. With instruments such as CONCERT on board ESA's Rosetta and Philae capable of probing the internal structure of large Solar System objects, studies focussing on the internal structure of the larger bodies, as determined by its growth and compaction history would be very interesting. The Monte Carlo method developed in this paper would be well suited for such studies, since adding parameters describing the aggregates is relatively straightforward.}


\subsection{Future work and implications}

\subsubsection{Pebble accretion}
\begin{figure}
\centering
\includegraphics[clip=,width=.9\linewidth]{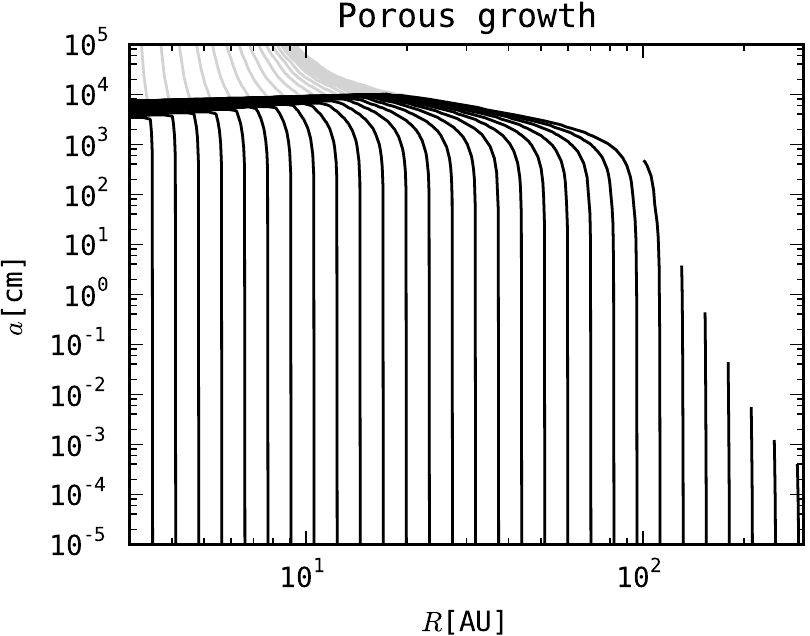}
\caption{\new{Same as the right-hand plot of Figure \ref{fig:semian2D}, but with aggregate size on the vertical axis.}}
\label{fig:semian2Db}
\end{figure}

A novel idea in the field of planet formation is the process of pebble accretion, where protoplanets grow very efficiently by accreting small pebbles \citep{ormelklahr2010,lambrechts2012, lambrechts2014, kretke2014}. These models rely on the radial influx of particles drifting in from the outer disk. As in the compact case, porous growth leads to the creation of rapidly drifting bodies in the region between 10 and $10^2$ AU (Figure \ref{fig:semian2D}). While the Stokes numbers of these particles are similar to the drifting pebbles in the compact case, their masses, sizes, and porosities can differ by many orders of magnitude \new{(see also Figure \ref{fig:semian2Db})}. In addition, the drag regime that the drifting bodies experience differs from the compact case (Figure \ref{fig:stokes}). Future studies are needed to address the effect of these factors on the efficiency of pebble accretion.

\subsubsection{Streaming instability}
While -- depending on the critical erosion velocity -- rapid coagulation into masses as large as planetesimals might be prevented by erosive collisions, the conditions created by this process might be favorable for triggering planetesimal formation by streaming instability \citep{youdin2005,johansen2007N,bai2010a,bai2010b}. To trigger streaming instability, the majority of mass needs to reside in particles with high Stokes numbers; the mid plane dust to gas ratio has to be close to unity; and the local vertically integrated dust-to-gas ratio needs to exceed ${\sim}0.03$ \citep{drazkowska2014b}. The first two conditions can be studied with simulations like the ones presented in this work. For example, for the steady-state distribution reached for $v_\mathrm{eros}=40\mathrm{~m~s^{-1}}$ at 5 AU for $\alpha=10^{-3}$, approximately $50\%$ of the dust mass resides in particles with $\Omega t_s > 10^{-2}$, and the mid plane dust-to-gas ratio is ${\sim}10^{-1}$. For weaker turbulence, the mid plane dust-to-gas ratio will be increased further, since $h_d \sim \alpha^{1/2}$ (Equation \ref{eq:h_d}). Because our simulations are local, the vertically integrated dust-to-gas ratio stays constant at $10^{-2}$. To fulfill the third condition, the dust-to-gas ratio either has to be larger from the beginning, or must increase by material drifting in from the outer disk. To study this, a global model is required, that calculates the evolution of the dust surface density in the presence of radial drift and erosion. In conclusion, drift-induced erosion appears to be a robust way of concentrating mass around $\Omega t_s \sim 1$, and is expected to create conditions favorable for streaming instability.

\subsubsection{The breakthrough case}
For compact silicate bodies in the inner disk, bouncing and fragmentation are very effective in stopping growth at mm-cm sizes. The breakthrough scenario, in which a small number of "lucky" particles still manages to gain mass, might render further growth possible \citep{windmark2012b, garaud2013}. The total mass fraction of these lucky particles can be extremely small, making this a challenging process to model for both differential and Monte Carlo methods \citep{drazkowska2014}. The distribution method used in this work, as outlined in Section \ref{sec:distribution_method}, is capable of resolving the entire mass distribution, including parts that contribute very little to the total dust mass, and appears to be well-suited for studying the breakthrough case. 

\subsubsection{Opacities of porous grains}
The optical properties of dust distributions resulting from porous growth are very different from populations containing exclusively solid particles. Not only are the mass distributions themselves different (e.g. Figure \ref{fig:semian2D}), but the scattering and absorption opacities of the individual grains are affected significantly by porosity \citep{kataoka2014,cuzzi2014}. For simple dust mass distributions, the effect of grain porosity on the appearance of protoplanetary disks has been investigated by \citet{kirchschlager2014}. Combining self-consistent coagulation models - including erosion and fragmentation - with porosity-dependent dust opacities will reveal the full impact porous growth has on the appearance of protoplanetary disks.

\section{Conclusions}\label{sec:conclusions}
Porous growth is very different from compact growth (Figure \ref{fig:semian2D}). For example, porous particles have larger collisional cross sections than compact particles of the same mass. More importantly, the aerodynamical properties of porous aggregates can differ greatly from those of compact particles (Figure \ref{fig:stokes}), causing differences in relative velocities (Figure \ref{fig:velocities}), vertical settling, and radial drift. 

We have modeled the coagulation of porous icy particles in the outer parts of protoplanetary disks, tracing the evolution of the mass and filling factor of the individual aggregates in time. We consider compaction through collisions, gas pressure, and self-gravity (Figure \ref{fig:internal_5}), and include a physical model for erosive collisions (Sections \ref{sec:erosion_case} and \ref{sec:erosion_model}). The main findings of this work are:
\begin{enumerate}

\item{Porous icy aggregates can outgrow the radial drift barrier in the inner ${\sim}10$ AU, despite increased growth timescales resulting from gas- and self-gravity compaction, if the perfect sticking assumption holds (Figures \ref{fig:drift_sticking_5_GG} and \ref{fig:semian2D}). This is in agreement with \citet{okuzumi2012} and \citet{kataoka2013a}.}

\item{While the maximum collision velocity between similar particles $({\sim}\alpha^{1/2}c_s)$ typically does not exceed the critical fragmentation threshold velocity for icy bodies, the velocity between drifting aggregates (with $\Omega t_s \geq 1$) and smaller bodies is much larger $({\sim} \eta v_\mathrm{K})$, and \new{can} exceed the critical threshold velocity for erosion (Figure \ref{fig:velocities}).}

\item{In these cases, we find that the mass loss through erosive collisions can balance the growth through same-size collisions, halting the growth of the largest bodies (Figures \ref{fig:projectile_E1} and \ref{fig:zeta}). In our local simulations, this results in a steady-state where the largest bodies have $\Omega t_s \sim 1$, and the porosity of the small fragment distribution is dominated by the fact that all fragments have at some point been part of these large eroded particles (Figures \ref{fig:m2fm_E1} and \ref{fig:fams}). Only for the highest erosion threshold velocity we considered ($v_\mathrm{eros}=60\mathrm{~m~s^{-1}}$), do the aggregates with $\Omega t_s\sim1$ manage to gain mass and grow through the drift barrier.}

\item{A simple semi-analytical model (Section \ref{sec:semian}) accurately describes the growth and drift behavior of the mass-dominating bodies. While no information is obtained about the dust mass distribution, such an approach is very useful for investigating how the size of the largest bodies depends on disk parameters such as the total disk mass, turbulence strength, or dust-to-gas ratio; and aggregate properties such as monomer size and erosion/fragmentation threshold velocities.}
\end{enumerate}

\begin{acknowledgements}
Dust studies at Leiden Observatory are supported through the Spinoza Premie of the Dutch science agency, NWO. S.K. would like to thank C.P. Dullemond, J. Dr{\c a}{\.z}kowska and T. Birnstiel for useful discussions, \new{and S. Okuzumi for comments and helping with the comparison to his work.} The authors thank S. Gundlach and J. Blum for sharing a version of their manuscript. C.W.O. acknowledges support for this work by NASA through Hubble Fellowship grant No. HST-HF-51294.01-A awarded by the Space Telescope Science Institute, which is operated by the Association of Universities for Research in Astronomy, Inc., for NASA, under contract NAS 5-26555.
\end{acknowledgements}

\bibliographystyle{aa}
\bibliography{refs}

\begin{thebibliography}{67}
\expandafter\ifx\csname natexlab\endcsname\relax\def\natexlab#1{#1}\fi

\bibitem[{{Bai} \& {Stone}(2010{\natexlab{a}})}]{bai2010a}
{Bai}, X.-N. \& {Stone}, J.~M. 2010{\natexlab{a}}, \apj, 722, 1437

\bibitem[{{Bai} \& {Stone}(2010{\natexlab{b}})}]{bai2010b}
{Bai}, X.-N. \& {Stone}, J.~M. 2010{\natexlab{b}}, \apjl, 722, L220

\bibitem[{{Birnstiel} {et~al.}(2010){Birnstiel}, {Dullemond}, \&
  {Brauer}}]{birnstiel2010}
{Birnstiel}, T., {Dullemond}, C.~P., \& {Brauer}, F. 2010, \aap, 513, A79

\bibitem[{{Blum} \& {Wurm}(2008)}]{blumwurm2008}
{Blum}, J. \& {Wurm}, G. 2008, \araa, 46, 21

\bibitem[{{Cuzzi} {et~al.}(2014){Cuzzi}, {Estrada}, \& {Davis}}]{cuzzi2014}
{Cuzzi}, J.~N., {Estrada}, P.~R., \& {Davis}, S.~S. 2014, \apjs, 210, 21

\bibitem[{{Dominik} \& {Dullemond}(2008)}]{dominikdullemond2008}
{Dominik}, C. \& {Dullemond}, C.~P. 2008, \aap, 491, 663

\bibitem[{{Dominik} \& {Tielens}(1995)}]{dominiktielens1995}
{Dominik}, C. \& {Tielens}, A.~G.~G.~M. 1995, Philosophical Magazine, Part A,
  72, 783

\bibitem[{{Dominik} \& {Tielens}(1997)}]{dominiktielens1997}
{Dominik}, C. \& {Tielens}, A.~G.~G.~M. 1997, \apj, 480, 647

\bibitem[{{Drazkowska} \& {Dullemond}(2014)}]{drazkowska2014b}
{Drazkowska}, J. \& {Dullemond}, C.~P. 2014, ArXiv e-prints

\bibitem[{{Dr{\c a}{\.z}kowska} {et~al.}(2014){Dr{\c a}{\.z}kowska},
  {Windmark}, \& {Dullemond}}]{drazkowska2014}
{Dr{\c a}{\.z}kowska}, J., {Windmark}, F., \& {Dullemond}, C.~P. 2014, \aap,
  567, A38

\bibitem[{{Dullemond} \& {Dominik}(2005)}]{dullemonddominik2005}
{Dullemond}, C.~P. \& {Dominik}, C. 2005, \aap, 434, 971

\bibitem[{{Garaud} {et~al.}(2013){Garaud}, {Meru}, {Galvagni}, \&
  {Olczak}}]{garaud2013}
{Garaud}, P., {Meru}, F., {Galvagni}, M., \& {Olczak}, C. 2013, \apj, 764, 146

\bibitem[{{Gillespie}(1975)}]{gillespie1975}
{Gillespie}, D.~T. 1975, Journal of Atmospheric Sciences, 32, 1977

\bibitem[{{Gundlach} {et~al.}(2011){Gundlach}, {Kilias}, {Beitz}, \&
  {Blum}}]{gundlach2011}
{Gundlach}, B., {Kilias}, S., {Beitz}, E., \& {Blum}, J. 2011, \icarus, 214,
  717

\bibitem[{{Gundlach} \& {Blum}(2014)}]{gundlach2014}
{Gundlach}, S. \& {Blum}, J. 2014, Accepted for publication in \apj

\bibitem[{{G{\"u}ttler} {et~al.}(2010){G{\"u}ttler}, {Blum}, {Zsom}, {Ormel},
  \& {Dullemond}}]{guttler2010}
{G{\"u}ttler}, C., {Blum}, J., {Zsom}, A., {Ormel}, C.~W., \& {Dullemond},
  C.~P. 2010, \aap, 513, A56

\bibitem[{{Haisch} {et~al.}(2001){Haisch}, {Lada}, \& {Lada}}]{haisch2001}
{Haisch}, Jr., K.~E., {Lada}, E.~A., \& {Lada}, C.~J. 2001, \apjl, 553, L153

\bibitem[{{Hayashi}(1981)}]{hayashi1981}
{Hayashi}, C. 1981, Progress of Theoretical Physics Supplement, 70, 35

\bibitem[{{Johansen} {et~al.}(2014){Johansen}, {Blum}, {Tanaka}, {Ormel},
  {Bizzarro}, \& {Rickman}}]{Johansen2014}
{Johansen}, A., {Blum}, J., {Tanaka}, H., {et~al.} 2014, ArXiv e-prints

\bibitem[{{Johansen} {et~al.}(2007){Johansen}, {Oishi}, {Mac Low}, {Klahr},
  {Henning}, \& {Youdin}}]{johansen2007N}
{Johansen}, A., {Oishi}, J.~S., {Mac Low}, M.-M., {et~al.} 2007, \nat, 448,
  1022

\bibitem[{{Kataoka} {et~al.}(2014){Kataoka}, {Okuzumi}, {Tanaka}, \&
  {Nomura}}]{kataoka2014}
{Kataoka}, A., {Okuzumi}, S., {Tanaka}, H., \& {Nomura}, H. 2014, \aap, 568,
  A42

\bibitem[{{Kataoka} {et~al.}(2013{\natexlab{a}}){Kataoka}, {Tanaka}, {Okuzumi},
  \& {Wada}}]{kataoka2013c}
{Kataoka}, A., {Tanaka}, H., {Okuzumi}, S., \& {Wada}, K. 2013{\natexlab{a}},
  \aap, 557, L4

\bibitem[{{Kataoka} {et~al.}(2013{\natexlab{b}}){Kataoka}, {Tanaka}, {Okuzumi},
  \& {Wada}}]{kataoka2013a}
{Kataoka}, A., {Tanaka}, H., {Okuzumi}, S., \& {Wada}, K. 2013{\natexlab{b}},
  \aap, 554, A4

\bibitem[{{Kempf} {et~al.}(1999){Kempf}, {Pfalzner}, \& {Henning}}]{kempf1999}
{Kempf}, S., {Pfalzner}, S., \& {Henning}, T.~K. 1999, \icarus, 141, 388

\bibitem[{{Kirchschlager} \& {Wolf}(2014)}]{kirchschlager2014}
{Kirchschlager}, F. \& {Wolf}, S. 2014, \aap, 568, A103

\bibitem[{{Kothe} {et~al.}(2010){Kothe}, {G{\"u}ttler}, \& {Blum}}]{kothe2010}
{Kothe}, S., {G{\"u}ttler}, C., \& {Blum}, J. 2010, \apj, 725, 1242

\bibitem[{{Kretke} \& {Levison}(2014)}]{kretke2014}
{Kretke}, K.~A. \& {Levison}, H.~F. 2014, ArXiv e-prints

\bibitem[{{Krijt} {et~al.}(2014){Krijt}, {Dominik}, \& {Tielens}}]{krijt2014}
{Krijt}, S., {Dominik}, C., \& {Tielens}, A.~G.~G.~M. 2014, Journal of Physics
  D Applied Physics, 47, 175302

\bibitem[{{Krijt} {et~al.}(2013){Krijt}, {G{\"u}ttler}, {Hei{\ss}elmann},
  {Dominik}, \& {Tielens}}]{krijt2013}
{Krijt}, S., {G{\"u}ttler}, C., {Hei{\ss}elmann}, D., {Dominik}, C., \&
  {Tielens}, A.~G.~G.~M. 2013, Journal of Physics D Applied Physics, 46, 5303

\bibitem[{{Lambrechts} \& {Johansen}(2012)}]{lambrechts2012}
{Lambrechts}, M. \& {Johansen}, A. 2012, \aap, 544, A32

\bibitem[{{Lambrechts} \& {Johansen}(2014)}]{lambrechts2014}
{Lambrechts}, M. \& {Johansen}, A. 2014, ArXiv e-prints

\bibitem[{{Meisner} {et~al.}(2012){Meisner}, {Wurm}, \& {Teiser}}]{meisner2012}
{Meisner}, T., {Wurm}, G., \& {Teiser}, J. 2012, \aap, 544, A138

\bibitem[{{Mukai} {et~al.}(1992){Mukai}, {Ishimoto}, {Kozasa}, {Blum}, \&
  {Greenberg}}]{mukai1992}
{Mukai}, T., {Ishimoto}, H., {Kozasa}, T., {Blum}, J., \& {Greenberg}, J.~M.
  1992, \aap, 262, 315

\bibitem[{{Nakagawa} {et~al.}(1986){Nakagawa}, {Sekiya}, \&
  {Hayashi}}]{nakagawa1986}
{Nakagawa}, Y., {Sekiya}, M., \& {Hayashi}, C. 1986, \icarus, 67, 375

\bibitem[{{Okuzumi} {et~al.}(2012){Okuzumi}, {Tanaka}, {Kobayashi}, \&
  {Wada}}]{okuzumi2012}
{Okuzumi}, S., {Tanaka}, H., {Kobayashi}, H., \& {Wada}, K. 2012, \apj, 752,
  106

\bibitem[{{Okuzumi} {et~al.}(2009){Okuzumi}, {Tanaka}, \&
  {Sakagami}}]{okuzumi2009}
{Okuzumi}, S., {Tanaka}, H., \& {Sakagami}, M.-a. 2009, \apj, 707, 1247

\bibitem[{{Okuzumi} {et~al.}(2011){Okuzumi}, {Tanaka}, {Takeuchi}, \&
  {Sakagami}}]{okuzumi2011}
{Okuzumi}, S., {Tanaka}, H., {Takeuchi}, T., \& {Sakagami}, M.-a. 2011, \apj,
  731, 95

\bibitem[{{Ormel} \& {Cuzzi}(2007)}]{ormel2007b}
{Ormel}, C.~W. \& {Cuzzi}, J.~N. 2007, \aap, 466, 413

\bibitem[{{Ormel} \& {Klahr}(2010)}]{ormelklahr2010}
{Ormel}, C.~W. \& {Klahr}, H.~H. 2010, \aap, 520, A43

\bibitem[{{Ormel} \& {Spaans}(2008)}]{ormel2008}
{Ormel}, C.~W. \& {Spaans}, M. 2008, \apj, 684, 1291

\bibitem[{{Ormel} {et~al.}(2007){Ormel}, {Spaans}, \& {Tielens}}]{ormel2007}
{Ormel}, C.~W., {Spaans}, M., \& {Tielens}, A.~G.~G.~M. 2007, \aap, 461, 215

\bibitem[{{Paszun} \& {Dominik}(2008)}]{paszun2008}
{Paszun}, D. \& {Dominik}, C. 2008, \aap, 484, 859

\bibitem[{{Paszun} \& {Dominik}(2009)}]{paszun2009}
{Paszun}, D. \& {Dominik}, C. 2009, \aap, 507, 1023

\bibitem[{{Poppe} {et~al.}(2000){Poppe}, {Blum}, \& {Henning}}]{poppe2000}
{Poppe}, T., {Blum}, J., \& {Henning}, T. 2000, \apj, 533, 454

\bibitem[{{Schr{\"a}pler} \& {Blum}(2011)}]{schrapler2011}
{Schr{\"a}pler}, R. \& {Blum}, J. 2011, \apj, 734, 108

\bibitem[{{Seizinger} \& {Kley}(2013)}]{seizinger2013a}
{Seizinger}, A. \& {Kley}, W. 2013, \aap, 551, A65

\bibitem[{{Seizinger} {et~al.}(2013){Seizinger}, {Krijt}, \&
  {Kley}}]{seizinger2013c}
{Seizinger}, A., {Krijt}, S., \& {Kley}, W. 2013, \aap, 560, A45

\bibitem[{{Sekiya} \& {Takeda}(2005)}]{sekiya2005}
{Sekiya}, M. \& {Takeda}, H. 2005, \icarus, 176, 220

\bibitem[{{Shakura} \& {Sunyaev}(1973)}]{shakura1973}
{Shakura}, N.~I. \& {Sunyaev}, R.~A. 1973, \aap, 24, 337

\bibitem[{{Suyama} {et~al.}(2008){Suyama}, {Wada}, \& {Tanaka}}]{suyama2008}
{Suyama}, T., {Wada}, K., \& {Tanaka}, H. 2008, \apj, 684, 1310

\bibitem[{{Suyama} {et~al.}(2012){Suyama}, {Wada}, {Tanaka}, \&
  {Okuzumi}}]{suyama2012}
{Suyama}, T., {Wada}, K., {Tanaka}, H., \& {Okuzumi}, S. 2012, \apj, 753, 115

\bibitem[{{Testi} {et~al.}(2014){Testi}, {Birnstiel}, {Ricci}, {Andrews},
  {Blum}, {Carpenter}, {Dominik}, {Isella}, {Natta}, {Williams}, \&
  {Wilner}}]{testi2014}
{Testi}, L., {Birnstiel}, T., {Ricci}, L., {et~al.} 2014, ArXiv e-prints

\bibitem[{{Tielens} {et~al.}(1994){Tielens}, {McKee}, {Seab}, \&
  {Hollenbach}}]{tielens1994}
{Tielens}, A.~G.~G.~M., {McKee}, C.~F., {Seab}, C.~G., \& {Hollenbach}, D.~J.
  1994, \apj, 431, 321

\bibitem[{{Wada} {et~al.}(2013){Wada}, {Tanaka}, {Okuzumi}, {Kobayashi},
  {Suyama}, {Kimura}, \& {Yamamoto}}]{wada2013}
{Wada}, K., {Tanaka}, H., {Okuzumi}, S., {et~al.} 2013, \aap, 559, A62

\bibitem[{{Wada} {et~al.}(2007){Wada}, {Tanaka}, {Suyama}, {Kimura}, \&
  {Yamamoto}}]{wada2007}
{Wada}, K., {Tanaka}, H., {Suyama}, T., {Kimura}, H., \& {Yamamoto}, T. 2007,
  \apj, 661, 320

\bibitem[{{Wada} {et~al.}(2009){Wada}, {Tanaka}, {Suyama}, {Kimura}, \&
  {Yamamoto}}]{wada2009}
{Wada}, K., {Tanaka}, H., {Suyama}, T., {Kimura}, H., \& {Yamamoto}, T. 2009,
  \apj, 702, 1490

\bibitem[{{Wada} {et~al.}(2011){Wada}, {Tanaka}, {Suyama}, {Kimura}, \&
  {Yamamoto}}]{wada2011}
{Wada}, K., {Tanaka}, H., {Suyama}, T., {Kimura}, H., \& {Yamamoto}, T. 2011,
  \apj, 737, 36

\bibitem[{{Weidenschilling}(1977)}]{weidenschilling1977}
{Weidenschilling}, S.~J. 1977, \mnras, 180, 57

\bibitem[{{Weidenschilling}(1980)}]{weidenschilling1980}
{Weidenschilling}, S.~J. 1980, \icarus, 44, 172

\bibitem[{{Whipple}(1972)}]{whipple1972}
{Whipple}, F.~L. 1972, in From Plasma to Planet, ed. A.~{Elvius}, 211

\bibitem[{{Windmark} {et~al.}(2012){Windmark}, {Birnstiel}, {Ormel}, \&
  {Dullemond}}]{windmark2012b}
{Windmark}, F., {Birnstiel}, T., {Ormel}, C.~W., \& {Dullemond}, C.~P. 2012,
  \aap, 544, L16

\bibitem[{{Wurm} {et~al.}(2004){Wurm}, {Paraskov}, \& {Krauss}}]{wurm2004}
{Wurm}, G., {Paraskov}, G., \& {Krauss}, O. 2004, \apj, 606, 983

\bibitem[{{Wurm} {et~al.}(2005){Wurm}, {Paraskov}, \& {Krauss}}]{wurm2005}
{Wurm}, G., {Paraskov}, G., \& {Krauss}, O. 2005, \icarus, 178, 253

\bibitem[{{Youdin} \& {Goodman}(2005)}]{youdin2005}
{Youdin}, A.~N. \& {Goodman}, J. 2005, \apj, 620, 459

\bibitem[{{Youdin} \& {Lithwick}(2007)}]{youdin2007}
{Youdin}, A.~N. \& {Lithwick}, Y. 2007, \icarus, 192, 588

\bibitem[{{Zsom} \& {Dullemond}(2008)}]{zsom2008}
{Zsom}, A. \& {Dullemond}, C.~P. 2008, \aap, 489, 931

\bibitem[{{Zsom} {et~al.}(2010){Zsom}, {Ormel}, {G{\"u}ttler}, {Blum}, \&
  {Dullemond}}]{zsom2010}
{Zsom}, A., {Ormel}, C.~W., {G{\"u}ttler}, C., {Blum}, J., \& {Dullemond},
  C.~P. 2010, \aap, 513, A57

\end{thebibliography}

\begin{appendix}
\end{appendix}

\end{document}